\documentclass[12pt]{article} 
\usepackage[margin=2.5cm]{geometry}
\usepackage{amssymb}
\usepackage{amsmath}
\usepackage{natbib}
\usepackage{subcaption}
\usepackage{url}
\usepackage{xcolor, hyperref}
\hypersetup{
    colorlinks,
    linkcolor={red!70!black},
    citecolor={blue!70!black},
    urlcolor={red!70!black}
}

\usepackage{times}
\usepackage{settings_custom}
\newcommand{\del}{\partial}
\newcommand{\Sest}{\boldsymbol{\hat{S}}}
\newcommand{\Sstar}{\bS^*}
\newcommand{\denshatnew}[1]{\hat{\sigma}({{#1}})}
\newcommand{\denshat}[1]{\hat{\sigma}_{{#1}}}
\newcommand{\dens}[1]{\sigma_{{#1}}}
\newcommand{\rotGroup}[1]{\mathcal{O}({#1})}
\newcommand{\Haar}[2]{\mu_{{#2}}({#1})}

\newcommand{\hrho}{{\hat{\rho}}}

\usepackage{mdframed}
\newmdtheoremenv{result}{Result}

\usepackage{graphicx}
\usepackage[capitalise]{cleveref}

\DeclareMathOperator{\MMSE}{MMSE} \DeclareMathOperator{\MSE}{MSE}

\begin{document}

\title{Optimal denoising of rotationally invariant rectangular matrices}

\author{Emanuele Troiani$^{1}$, Vittorio Erba$^1$, \\ Florent Krzakala$^2$, Antoine Maillard$^3$, Lenka Zdeborová$^1$}

\date{
    \small
    $^1$Statistical Physics of Computation Lab, \'Ecole Polytechnique F\'ed\'erale de Lausanne (EPFL)\\
    $^2$Information, Learning and Physics lab, \'Ecole Polytechnique F\'ed\'erale de Lausanne (EPFL)\\
    $^3$Department of Mathematics \& Institute for Mathematical Research (FIM), ETH Zurich
}
\maketitle

\begin{abstract}%
    In this manuscript we consider denoising of large rectangular matrices:
    given a noisy observation of a signal matrix, what is the best way of
    recovering the signal matrix itself? For Gaussian noise and
    rotationally-invariant signal priors, we completely characterize the optimal
    denoiser and its performance in the high-dimensional limit, in which the
    size of the signal matrix goes to infinity with fixed aspects ratio, and
    under the Bayes optimal setting, that is when the statistician knows how the
    signal and the observations were generated. Our results generalise previous
    works that considered only symmetric matrices to the more general case of
    non-symmetric and rectangular ones. We explore analytically and numerically
    a particular choice of factorized signal prior that models cross-covariance
    matrices and the matrix factorization problem. As a byproduct of our
    analysis, we provide an explicit asymptotic evaluation of the rectangular
    Harish-Chandra-Itzykson-Zuber integral in a special case.
\end{abstract}


\section{Introduction}

In this paper we consider the problem of denoising large rectangular matrices,
i.e.\ the problem of reconstructing a matrix $\Sstar \in \bbR^{m \times p}$ from
a noisy observation $\bY = P_{\rm noise}(\Sstar)  \in \bbR^{m \times p}$. Our
aim is to characterize theoretically this problem in the Bayes optimal setting,
in which the statistician knows the details of the prior distribution of the
signal $P_{\rm signal}$ and the noisy channel $P_{\rm noise}$. In particular, we
will consider the case of additive Gaussian noise $P_{\rm noise}(\Sstar) =
\Sstar + \sqrt{\Delta} \bZ$ where $\Delta > 0$ controls the strength of the
noise, and $\bZ \in \bbR^{m \times p}$ is a matrix of i.i.d.\ Gaussian random
variables. On the side of the signal we will consider rotationally invariant
priors, i.e.\ those where $P_{\rm signal}(\bS^*) = P_{\rm signal} (\bU \bS^*
\bV)$ for any pair of rotation matrices $\bU \in \rotGroup{m}, \bV \in
\rotGroup{p}$.

We are interested in particular in the case of factorized priors, i.e.\ $\Sstar
= \bF \bX$ with $\bF \in \bbR^{m \times r}$ and $\bX \in \bbR^{r \times p}$, and
both $\bX$ and $\bF$ having i.i.d.\ Gaussian entries. From the point of view of
applications, this is equivalent to the problem of denoising cross-correlation
matrices, which is, for example, extremely relevant in modern financial
portfolio theory where it allows to compute better performing investment
portfolios by reducing overfitting \cite{cleaning2017}.

From a theoretical point of view, the denoising problem is a simpler variant of
the matrix factorization problem, where one would like to reconstruct the two
factors $\bF$ and $\bX$ from the noisy observation $\bY$. Matrix factorization
is ubiquitous in computer science, modeling many applications, from sparse PCA
\cite{sparsePCA} to dictionary learning \cite{online_dictionary}. While well
studied in the low-rank regime $r = \Theta(1)$ and  $\min(p, m) \to \infty$,
where optimal estimators and guarantees on their performances are available
\cite{2017Thibault,miolane2018fundamental}, much less is known in the
extensive-rank regime, where $r$ is of the same order as $m$ and $p$. 

This setting was studied, for generic priors on elements of $\bF$ and $\bX$, in
\cite{kabashima}, where an analysis of the information-theoretic thresholds and
the performance of approximate message passing algorithms was given. A more
recent series of works pointed out however that a key assumption of
\cite{kabashima} does not hold in practice, and proposed alternative analysis
techniques, ranging from spectral characterizations
\cite{schmidt:tel-03227132,barbier2021statistical} to high-temperature
expansions \cite{maillard2021perturbative}. In all these cases, the analytical
results obtained for rectangular matrices are not explicit and of limited
practical applicability even in the very simple case of Gaussian priors on the
factors and additive Gaussian noise. Studying the denoising problem in the
Gaussian regime is thus a first step towards a better understanding of matrix
factorization in this challenging regime.

Our main results concern \textit{rotationally invariant} priors, i.e.\ signal
matrices whose information lies exclusively in the distribution of their
singular values, corrupted by additive Gaussian noise. For this large class of
priors, we compute the optimal denoiser (in the sense of the mean square error
on the matrix $\bS$) and its performance in the Bayes optimal setting, and we
discuss in detail the case of factorized priors where both the factors $\bF$ and
$\bX$ have i.i.d.\ Gaussian components. As a byproduct of our analysis, we also
provide an explicit formula for the high-dimensional asymptotics of the
rectangular HCIZ integral \cite{HC,IZ,guionnet2021large} in a special case.

The code for the numerical simulations and for reproducing the figures is
available here: \url{https://github.com/PenombraET/rectangular_RIE}

\subsection{Definition of the model}\label{sec.def}

In this paper we focus our attention to rotationally-invariant priors, i.e.
$P_{\rm signal}(\bS) = P_{\rm signal} (\bU \bS \bV)$ for any pair of rotation
matrices $\bU \in \rotGroup{m}, \bV \in \rotGroup{p}$, where $\rotGroup{m}$ is
the orthogonal group in $m$ dimensions, and additive white Gaussian noise $\bZ$
being a matrix of i.i.d.\ Gaussian random variables with zero mean and variance
$(mp)^{-1/2}$\footnote{ We decided to use a \textit{symmetric normalization},
that is to scale all quantities by $\sqrt{mp}$ instead of using $m$ or $p$,
highlighting the  symmetry under transposition of the problem. In matrix
factorization applications, $m$ would typically denote the number of samples,
$p$ the dimensionality of the samples, and a more common normalization
convention would require to normalize all quantities using $p$. Our results can
be adapted accordingly, as this normalization change amounts to an overall
rescaling by $\sqrt{R_1}$. }. The rotational invariance of the prior, together
with the rotational invariance of Gaussian noise, implies that the observation
$\bY$ will have a rotationally-invariant distribution, and that all relevant
observables of the problem will depend only on the singular value distributions
of $\bS, \bZ$ and $\bY$. 

In order for the signal-to-noise ratio (SNR)  $\sqrt{\Delta}$ to be comparable
over different choices of priors, we fix the ratio between the averaged $L^2$
norm of the signal matrix $\bS$ and that of the noise matrix $\bZ$ to 1.
Explicitly, one requires that 
\begin{equation}
  \mathbb{E}_{\rm prior}[||\bS||_2^2] = \EE_{\rm noise}[||\bZ||_2^2 ] = \sqrt{mp} \, ,
\end{equation}
where the $L^2$ norm is defined as $||\bS||_2^2 = \Tr(\bS \bS^T)$. We will
consider the high-dimensional regime, i.e.\ the limit $m, p \to \infty$ with
fixed ratio $R_1 \equiv p/m.$ Without loss of generality, we will consider $p
\geq m$, i.e.\ $R_1 \geq 1$. We require that in this limit the singular values
of the signal are of order $\mathcal{O}(1)$, and that the corresponding
empirical singular value density of the prior converges to a deterministic
probability density function.

As a particular choice of rotationally-invariant prior, we focus on on
factorized signals, i.e.
\begin{equation}\label{eq.defWishart}
  \Sstar = \frac{\bF \bX}{\sqrt{r} \sqrt[4]{m p}} \, ,
\end{equation}
where $\bF \in \mathbb{R}^{m \times r}$, $\bX \in \mathbb{R}^{r \times p}$ are
matrices with i.i.d.\ standard Gaussian entries. In the high dimensional limit,
we will consider the extensive-rank regime, where $R_2 \equiv r/m$ is kept
constant. We will also study the low-rank limit of this prior, i.e. the limit
$R_2 \to 0$, or equivalently $r \ll m$. This form of the prior models both
Gaussian-factors matrix factorization and cross-covariance matrices of two
datasets of $r$ samples in dimensions respectively $m$ and $p$.

\subsection{Main Results}\label{sec.results}

Our main result is the analytical characterization of the optimal denoiser and
its predicted performance in the high-dimensional Bayes-optimal setting, i.e.\
when the statistician knows the details of the prior distribution of the signal
and the noisy channel, for rotationally-invariant priors and additive Gaussian
noise. The optimal denoiser here is defined as the function $\bY \mapsto
\Sest(\bY)$ of the observation that minimizes the average mean-square error
(MSE) with the ground truth,
\begin{equation}
  \Sest(\cdot) = \argmin_{\text{denoisers}\,\, f} \frac{1}{\sqrt{mp}}\EE_{\Sstar, \bY} ||\Sstar - f(\bY) ||_2^2 \, ,
\end{equation}
where $\EE_{\Sstar, \bY}$ is the joint average over the ground truth and the
noisy observation. Similarly we define the minimal mean-square error (MMSE) as
the averaged MSE of the optimal estimator:

\begin{equation}\label{eq.defMMSE}
    \MMSE = \EE_{\Sstar, \bY}\left[ \MSE\left(\Sstar, 
    \Sest(\bY) \right) \right] \, .
\end{equation}

In order to state our results, let us denote by $\denshat{\bY}$ the symmetrized
asymptotic singular value density of a $\bY$-distributed matrix, i.e.
\begin{equation}
    \denshat{\bY}(x) = 
    \lim_{m \to \infty} \frac{1}{m} \sum_{i = 1}^{m} 
    \left[ \frac{1}{2} \delta(x - y_i) + \frac{1}{2} \delta(x+ y_i) \right],
\end{equation}
where $y_i$ are the singular values of a $\bY$-distributed matrix of size $m
\times R_1 m$. Let us also denote by $\bY = \bU \bLambda \bV$ the singular value
decomposition (SVD) of the actual instance of the observation $\bY$, where $\bU
\in \bbR^{m \times m}$ is orthogonal by rows, $\bV \in \bbR^{m \times p}$ is
orthogonal by columns and $\bLambda = \diag(\lambda_1, \dots, \lambda_m)$ is the
diagonal matrix of singular values of $\bY$\footnote{By concentration of
spectral densities of large matrices, the empirical spectral density of an
actual instance of $\bY$ converges to the deterministic asymptotic spectral
density $\denshat{\bY}$ in the high-dimensional limit.}.

\begin{result}[Optimal estimator]\label{res.est} The optimal estimator is
rotationally-invariant, i.e.\ diagonal in the basis of singular vectors of the
observation $\bY$, and it is given by 
\begin{equation}\label{eq.res1}
  \Sest(\bY) = \bU \diag( \xi(\lambda_1), \dots, \xi(\lambda_m)  ) \bV,
\end{equation}
where the spectral denoising function $\xi$ is given, in the limit $m,p \to
\infty$ with $R_1 = p / m$ fixed, by 
\begin{equation}
  \xi(\lambda) = \lambda - \frac{2 \Delta}{\sqrt{R_1}} \left[ \frac{R_1 -1}{2\lambda} + \dashint d\zeta\, 
  \frac{\denshat{\bY}(\zeta)}{\lambda-\zeta} \right], \label{eq:rescaling}
\end{equation}
and the integral is intended as a Cauchy principal value integral.
\end{result}
We obtain \cref{res.est} in \cref{sec.optimal} by computing the average of the
posterior distribution $P(\bS\mid\bY)$, i.e.\ the probability that a candidate
signal $\bS$ was used to generate the observation $\bY$, 
\begin{equation}\label{eq.posterior3}
  \begin{split}
    P(\bS \mid \bY) 
    &= \frac{1}{\caZ_{\bY}} P_{\rm signal}(\bS) \Delta^{-\frac{mp}{2}} \exp\left[-\frac{\sqrt{mp}}{2\Delta}
    \Tr\left( ( \bY- \bS) ( \bY- \bS)^T  \right)\right]
    \, ,
  \end{split}
\end{equation}
where $\caZ_{\bY}$ is the partition function, i.e.\ the correct normalization
factor
\begin{equation}\label{eq.partition3}
  \begin{split}
    \caZ_{\bY}
    &= \int d\bS \, P_{\rm signal}(\bS) \Delta^{-\frac{mp}{2}} \exp\left[-\frac{\sqrt{mp}}{2\Delta}
    \Tr\left( ( \bY- \bS) ( \bY- \bS)^T  \right)\right]
    \, .
  \end{split}
\end{equation}
Indeed, one can prove that the posterior average is always the optimal estimator
with respect to the MSE metric \cite{cover1999elements}.

\begin{result}[Analytical MMSE]\label{res.mmse} In the limit $m,p \to \infty$
 with $R_1 = p / m$ fixed, the MMSE is given by
\begin{equation}\label{eq.res2}
  \begin{split}
      \MMSE
      = \Delta - 2 \Delta^2 \frac{\del}{\del \Delta} 
      \Bigg[ & \frac{1}{R_1}  \dashint d\lambda \, d\zeta \,  \denshat{\bY}(\lambda) \denshat{\bY}(\zeta) \log|\lambda-\zeta|  \\ 
      &\quad+ \frac{R_1 - 1}{R_1}  \dashint d\lambda \, \denshat{\bY}(\lambda) \log|\lambda| \Bigg]
  \, ,
  \end{split}
\end{equation}
where the dashed integral signs denote the symmetric regularization of the
integrals (\textit{\'a la} Cauchy principal value) around the singularities of
the logarithms --- see \cref{app.regularization}. 
\end{result}
We obtain \cref{res.mmse} in \cref{sec.optimal} by using the I-MMSE theorem
\cite{guo2004mutual}, which links the performance of optimal estimators in
problems with Gaussian noise with the derivative of the partition function
$\caZ_{\bY}$ with respect to the SNR.

Notice that to implement numerically the denoising function \cref{eq.res1} and
the MMSE \cref{eq.res2} one needs to compute the symmetrized asymptotic singular
value density of the observation $\bY$, $\denshat{\bY}$. We will provide details
on how to compute it for generic rotationally-invariant priors in
\cref{sec.general} and in the special case of the Gaussian factorized prior in
\cref{sec.factorized}.

On a more technical note, the computation of the partition function
\cref{eq.partition3}, from which \cref{res.est} and \cref{res.mmse} are derived,
involves the computation of the asymptotics of a rectangular
Harish-Chandra-Itzykson-Zuber (HCIZ) integral, defined as
\begin{equation}\label{eq.HCIZdef}
  \caI_{m}(\bA, \bB; \tau) = 
  \int \Haar{d \bU}{m} \Haar{d \bV}{p}
  \exp\left[\tau m
  \Tr\left(   \bA \bV \bB^T \bU \right)\right]
\end{equation}
for any pair of rectangular matrices $\bA, \bB \in \bbR^{m \times p}$ and $\tau
> 0$, where $\Haar{\cdot}{m}$ is the uniform measure over the $m$-dimensional
orthogonal group $\rotGroup{m}$. Notice that the HCIZ integral depends only on
the singular values of its arguments, as the singular vectors can always be
reabsorbed in the integration over orthogonal groups.

We will justify in detail why the HCIZ integral appears in the computation of
\cref{eq.partition3} in \cref{sec.optimal}. The idea is that, after a change of
variables to the SVD decomposition of $\bS$ in \cref{eq.partition3}, the
coupling term $\Tr (\bS \bY^T)$ will be the only term depending on the singular
vectors of $\bS$. This term, together with integration over the singular vectors
of $\bS$, will give rise to a rectangular HCIZ integral $\caI_{m}(\bS, \bY;
\sqrt{R_1}/\Delta)$. The integration over the spectrum of $\bS$ will be
performed explicitly thanks to a combination of a concentration argument and
Nishimori identities \cite{Nishimori_1980}, so that the actual HCIZ integral we
will be interested in is $\caI_{m}(\Sstar, \bY; \sqrt{R_1}/\Delta)$.


In general, the asymptotics of the HCIZ integral is difficult to characterize in
closed form, and it is linked to the solution of a 1d hydrodynamical problem
\cite{Matytsin, guionnet2021large}. In the special case in which the two
matrices over which the HCIZ integral is evaluated differ only by a matrix of
i.i.d.\ Gaussian variables --- which happens to be the case in our computation,
as $\bY - \Sstar = \sqrt{\Delta} \bZ$ --- this non-trivial problem simplifies,
allowing for a closed from computation of the asymptotics of the HCIZ integral. 

\begin{result}[Asymptotics of the rectangular HCIZ integral]\label{res.hciz}
 Under the hypotheses of \cite[Theorem 1.1]{guionnet2021large}, and in the case
 in which $\bB = \bA + \kappa \bZ$ with $\bZ$ a matrix of i.i.d.\ Gaussian
 variables with zero mean and variance $m^{-1}$, one has
\begin{equation}\label{eq.HCIZres}
  \begin{split}
    &I_{R_1}\left[\denshat{\bA}, \denshat{\bB}; \tau \right] =
    \lim_{\substack{\\m \to \infty\\ p = R_1 m}}
    \frac{2}{m^2} \log \caI_{m}\left(\bA, \bB ; \tau \right)
    \\
    &\quad= 
    C(R_1, \kappa \sqrt\tau) + R_1 \log\Delta
    + \frac{\sqrt{R_1}}{\Delta} \int d\lambda \, \denshat{\bA}(\lambda) \lambda^2
    + \frac{\sqrt{R_1}}{\Delta} \int d\lambda \, \denshat{\bB}(\lambda) \lambda^2
    \\ &\qquad
    - 2 (R_1 - 1) \dashint d\lambda \, \denshat{\bB}(\lambda) \log|\lambda|
    - 2 \dashint d\lambda \, d\zeta \, \denshat{\bB}(\lambda)  \denshat{\bB}(\zeta) \log|\lambda - \zeta|,
  \end{split}
\end{equation}
where $\denshat{\bA, \bB}$ is the symmetrized asymptotic singular value density
of, respectively, $\bA$ or $\bB$ and $C(R_1, \kappa \sqrt\tau)$ is an
undetermined constant depending only on $R_1$ and on the product $\kappa
\sqrt\tau$. Again, dashed integrals need to be regularized as detailed in
\cref{app.regularization}. 
\end{result}

We justify \cref{res.hciz} in \cref{sec.HCIZ} by specifying the general
asymptotic form given in \cite{guionnet2021large} to the $\bB = \bA + \kappa
\bZ$ case. This result generalizes Result 3.2 of \cite{maillard2021perturbative}
for the case of symmetric matrices to the case of rectangular ones.

Finally, let us note that we stated the main findings of the paper as \textit{Results} instead of \textit{Theorems} to highlight that we shall not provide a complete rigorous justification for each steps of computation, and leave some technical details to the reader. Nonetheless, we believe that each of our derivation could be made entirely rigorous through a slightly more careful control.

\subsection{Numerical results and comparisons}\label{sec.numerics}

Before presenting the technical details that justify \cref{res.est},
\cref{res.mmse} and \cref{res.hciz}, we provide numerical simulations in the
case of the Gaussian factorized prior \cref{eq.defWishart}. We have two aims:
(i) corroborating our analytical result by showing that on actual instances of
the denoising problem the performance of our estimator \cref{eq.res1} (empirical
MMSE) equals that predicted by the MMSE formula \cref{eq.res2} (analytical
MMSE); (ii) studying the phenomenology of the MMSE as a function of the noise
level $\Delta$, the aspect-ratio $R_1$ and the rank-related parameter $R_2$.

To implement numerically the denoising function \cref{eq.res1} and the MMSE
\cref{eq.res2} one needs to compute the symmetrized singular value density of
the observation $\bY$, $\denshat{\bY}$. We will provide details on how to
compute it in the special case of the Gaussian factorized prior in
\cref{sec.factorized}. Given $\denshat{\bY}$, one can compute (i) the empirical
MMSE by considering a random instance of the observation $\bY = \Sstar +
\sqrt{\Delta} \bZ$, by denoising it with the optimal denoiser $\Sest(\cdot)$,
and by computing the MSE between $\Sstar$ and the cleaned matrix $\Sest(\bY)$;
(ii) the analytical MMSE \cref{eq.res2} by numerical integration of
$\denshat{\bY}$. Details on numerical integration are given in
\cref{sec.tricks}. At fixed $R_1$, we distinguish two regimes for $R_2$:
over-complete $R_2 > 1$ and under-complete $R_2 < 1$. 

\begin{figure}
\centering
\includegraphics[width=0.49\columnwidth]{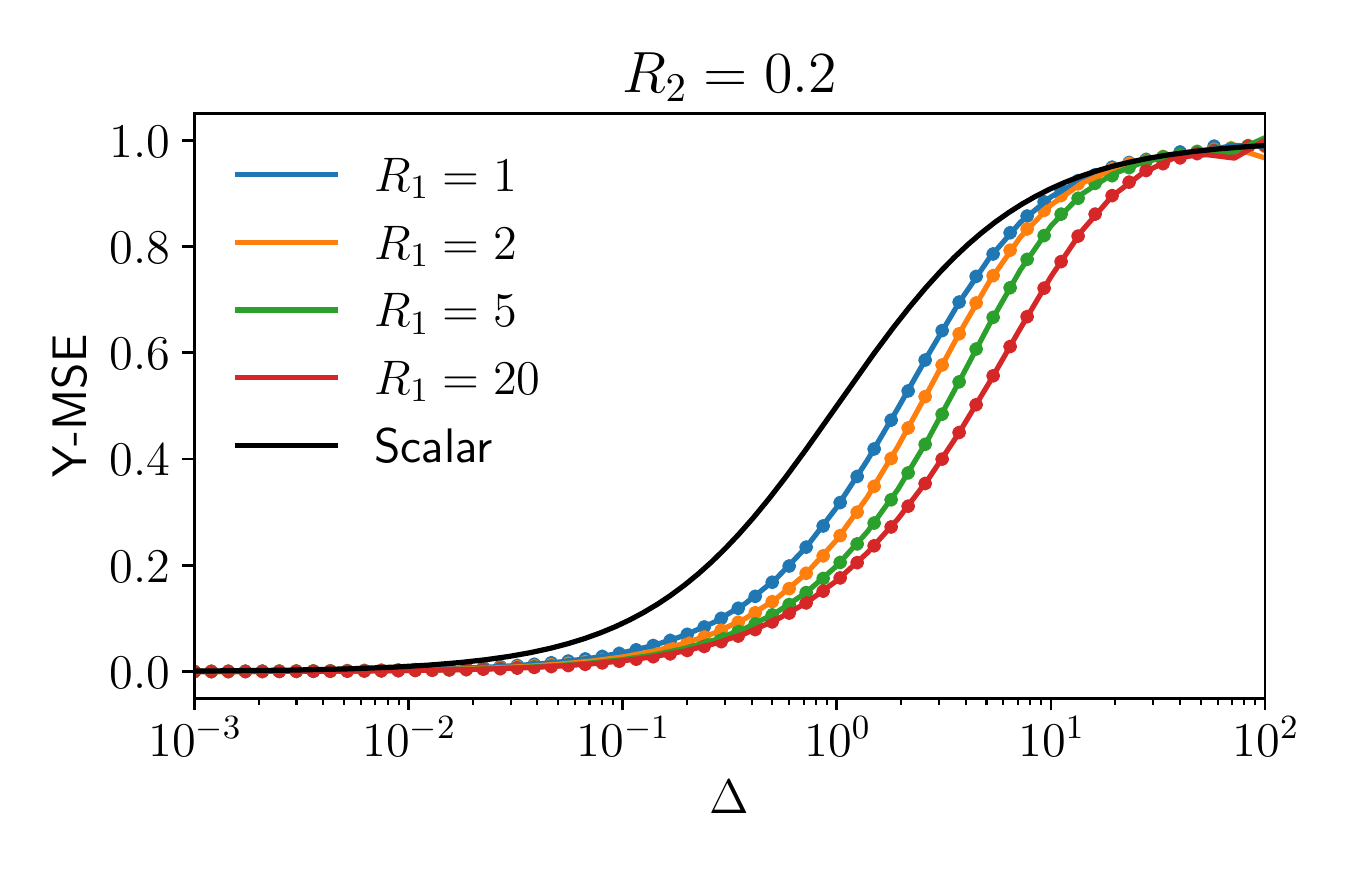}
\includegraphics[width=0.49\columnwidth]{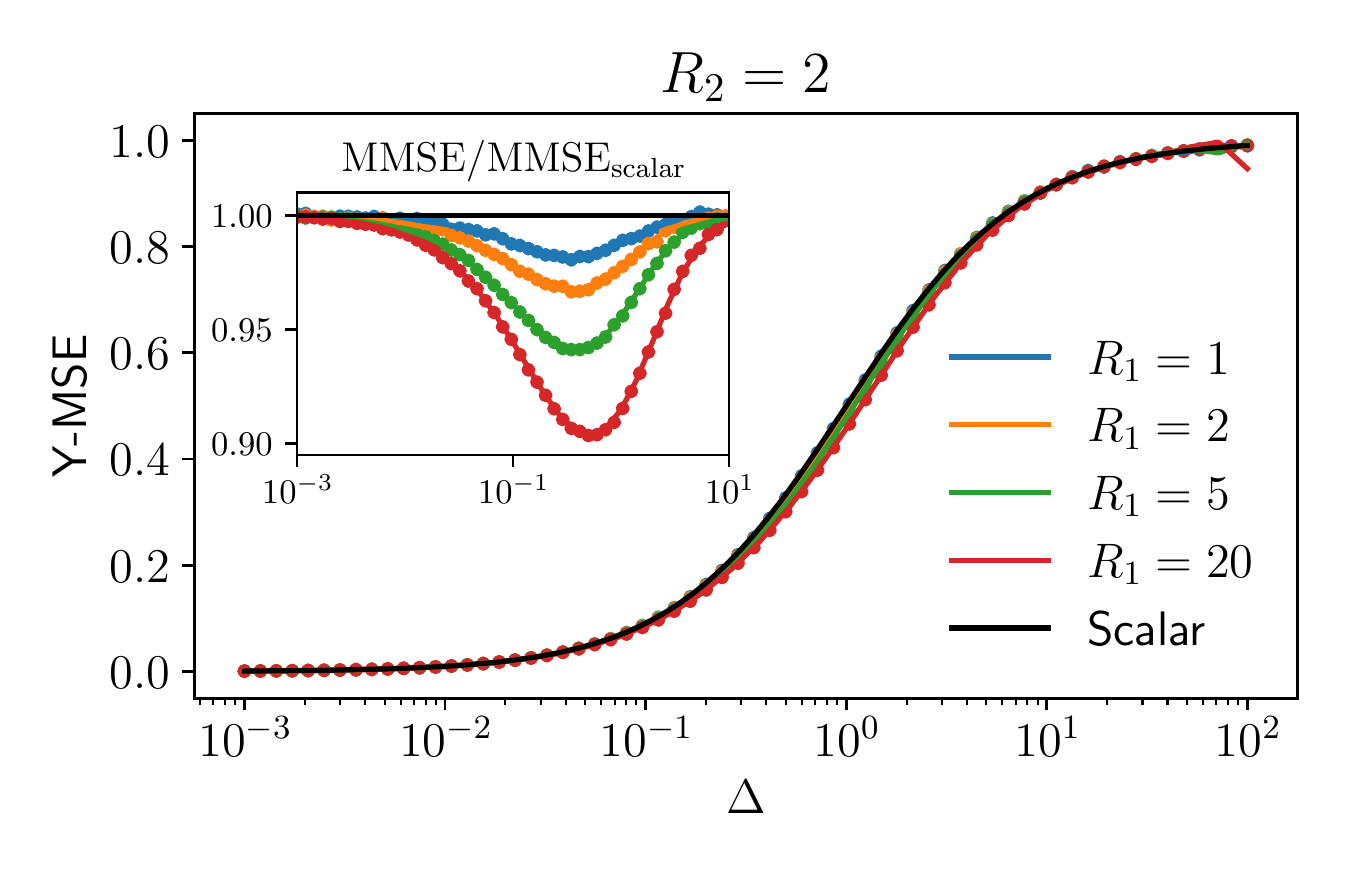}
\vspace{-5mm}
\caption{
Overview of the behaviour of the MMSE for the Gaussian factorized priors. In each plot, colored lines denote the analytical MMSE \cref{eq.res2}, colored dots the empirical MMSE estimated on single instances at size $m=2000$ while the black line shows the MMSE of scalar denoising. In all panels we observe a perfect agreement between theory and simulations.
The left panel shows the dependence of the MMSE on $R_1$ in the under-complete case $R_2 < 1$: we observe that the MMSE greatly improves as $R_1$ grows. 
The right shows the same in the over-complete case: the dependence on $R_1$ in this regime is very small, and it is highlighted in the inset where the y axis has been rescaled using the $R_2 \to \infty$ limit of the MMSE, that of scalar denoising. \vspace{-5mm}
}
\label{fig_1}
\end{figure}

In the under-complete regime $R_2 < 1$, the rank of the signal matrix is
non-maximal. \cref{fig_1} shows a strong dependence of the MMSE on $R_1$ for
$R_2 <1$, with better MMSE for larger $R_1$. In particular, we observe that the
MMSE at a given value of $\Delta$ decreases as $R_1$ grows larger, and the
cross-over between low and high denoising error shifts to larger values of
$\Delta$ (lower SNR). This is in accordance with intuition: larger $R_1$
correspond to matrices with higher aspect-ratio, and thus with a larger number
of components (recall that $\bY$ is an $m \times R_1 m$ matrix), while the (low)
rank $R_2 m$ --- here measuring the inverse degree of correlations between
components of $\bY$ --- remains fixed.

\begin{figure} \centering \begin{minipage}[c]{0.49\columnwidth} \caption{
Limiting behaviour of the MMSE for $R_2 \to \infty$. At fixed $R_1$, the plot
show convergence to the scalar denoising limit as $R_2 \to \infty$. Colored
lines denote the analytical MMSE \cref{eq.res2}, colored dots the empirical
MMSE estimated on single instances at size $m=2000$ and black lines denote the
scalar denoising $\MMSE_{\rm scalar} = \Delta / (\Delta +1)$.
}
\label{fig_high_rank} \end{minipage} \begin{minipage}[c]{0.49\columnwidth}
\includegraphics[width=\columnwidth]{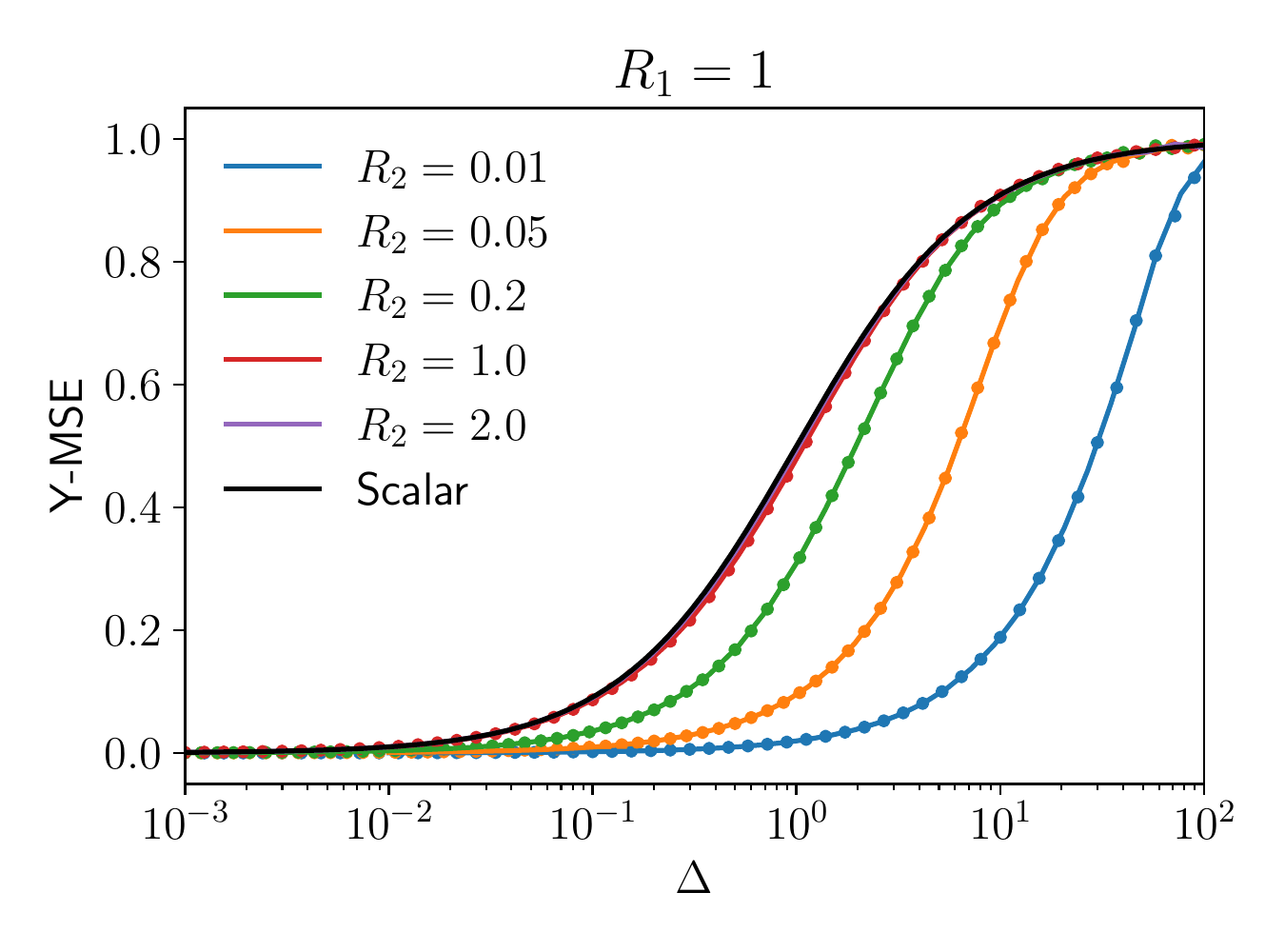} \end{minipage} \vspace{-5mm} \end{figure}

In the over-complete regime $R_2 > 1$, and in particular for $R_2 \to \infty$,
the prior trivializes: each of the components of $\bY$ is a sum of a large
enough number of independent variables for the central limit theorem to hold.
Thus, $\Sstar$ becomes a matrix with i.i.d.\ coordinates, and our problem
factorizes into simpler scalar denoising problems on each of the matrix
components. \cref{fig_1} portrays the over-complete regime for $R_2 = 2$. We
see that the dependence on $R_1$ is extremely weak --- see inset in
\cref{fig_1} --- signaling a very fast convergence towards scalar denoising.
In the $R_2 \to \infty$ limit --- see \cref{fig_high_rank} --- the MMSE
converges to that of scalar denoising, $\MMSE_{\rm scalar} = \Delta /
(1+\Delta)$, independently on $R_1$.

\begin{figure}
\centering
\includegraphics[width=0.475\columnwidth]{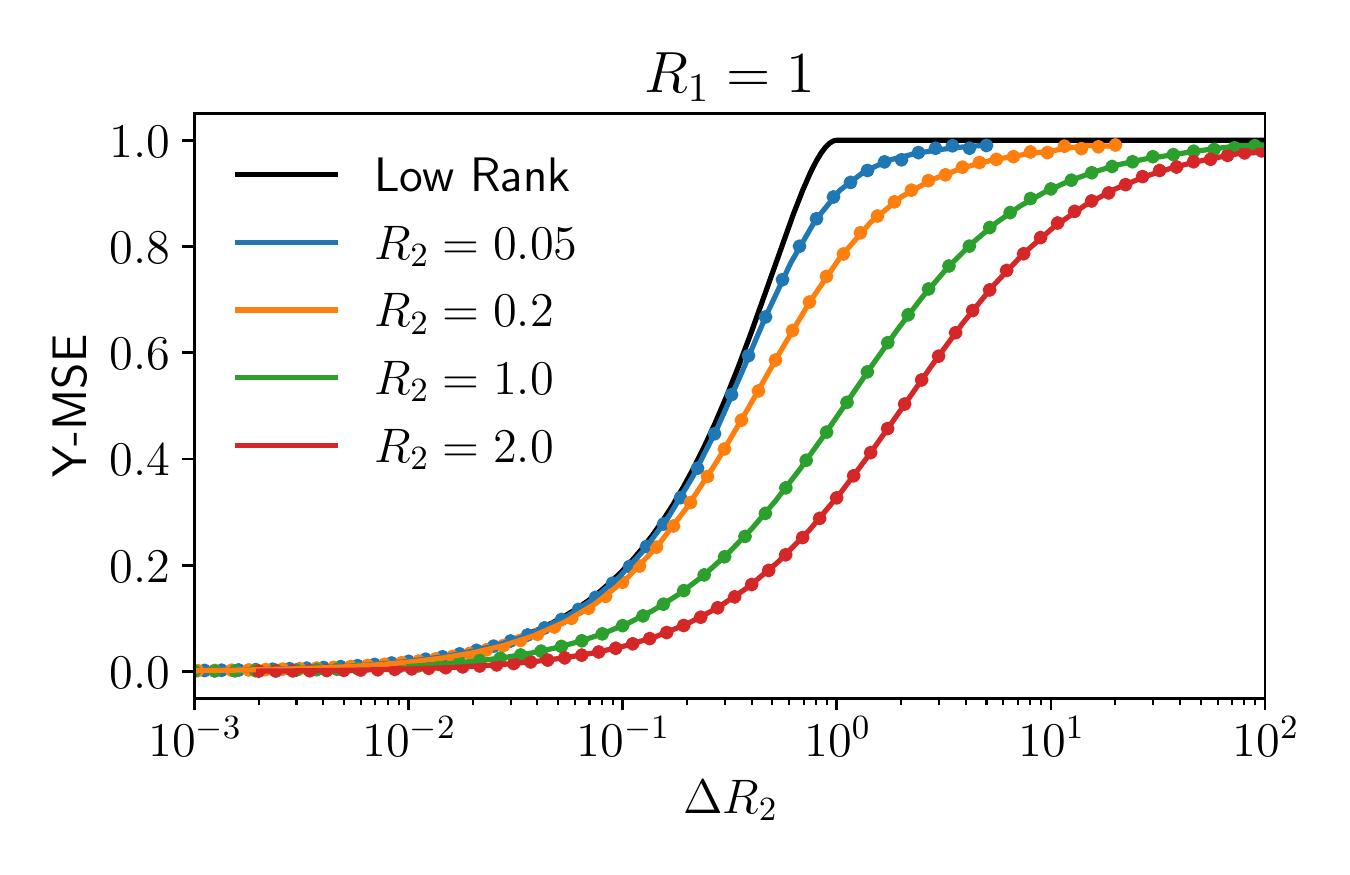}
\includegraphics[width=0.475\columnwidth]{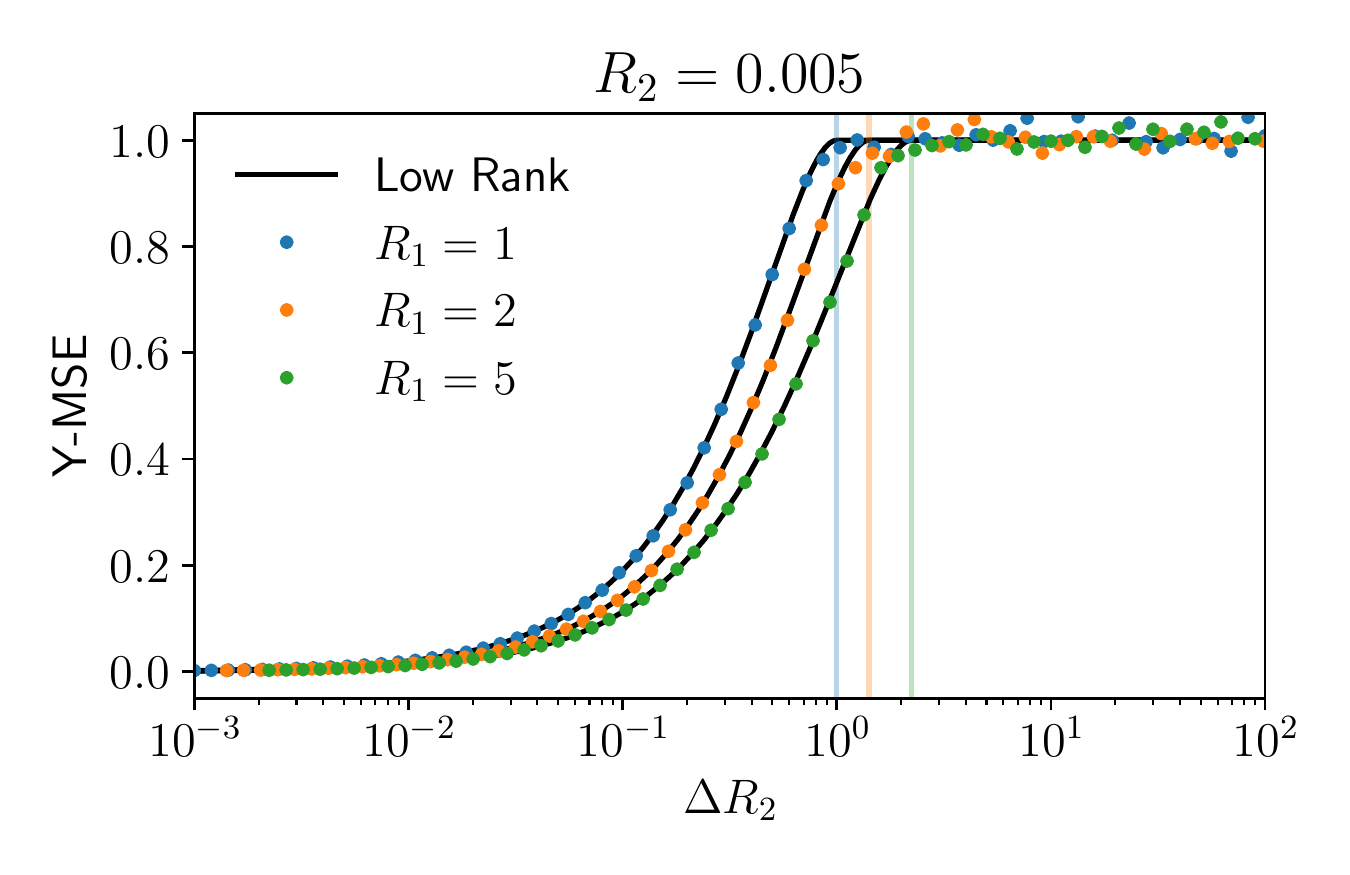}
\vspace{-5mm}
\caption{
Limiting behaviour of the MMSE for $R_2 \to 0$.
In each plot, colored lines denote the analytical MMSE \cref{eq.res2}, colored dots the empirical MMSE estimated on single instances at size $m=2000$ and black lines denote the low-rank limiting behaviour. We rescaled the horizontal 
axis to better highlight convergence to the limit.
The left panel shows, at fixed $R_1$, convergence to the low-rank limit as $R_2 \to 0$.
The right shows, at fixed small $R_2$, the dependence on $R_1$ of the low-rank behaviour. Vertical lines highlight the critical value $R_2 \Delta_c = \sqrt{R_1}$ at which the MMSE changes behaviour: above this threshold, no denoising is possible at low-rank.\vspace{-5mm}
}
\label{fig_low_rank}
\end{figure}

In the extreme low-rank limit $R_2 \to 0$, we expect to recover the results of
low-rank matrix denoising \cite{baik2004phase,2017Thibault} as already shown for
the symmetric-matrix denoising in \cite{maillard2021perturbative}. In
particular, we expect to recover the MMSE phase transition at $\Delta_c =
\sqrt{R_1} / R_2$. \cref{fig_low_rank} shows convergence towards the phase
transition for a fixed value of $R_1$, while \cref{fig_low_rank} confirms that
the dependence of $\Delta_c$ on $R_1$ is the expected one. We give details on
how to compute the theoretical MMSE in the low-rank limit in \cref{app.lowrank}.

\subsection{Related works}\label{sec.related} 

Spectral denoisers for covariance matrices have been used extensively in
finance, as in \cite{2012Ledoit, RIEoriginal, cleaning2017}. Very recently
\cite{benaychgeorges2021optimal} proposed an algorithm to denoise cross
correlation matrices that is in spirit similar to the one we propose, with a
crucial difference. In our setting, the signal matrix $\Sstar = \bF \bX$ is
corrupted by a noisy channel, and the objective is to get rid of the noise. In
their setting, they are given two correlated datasets $\bF$ and $\bX$, and the
objective is to estimate their theoretical cross-covariance, which is a
non-trivial task when the number of observations $r$ is comparable with the
dimensions $m, p$. Qualitatively, in their setting the noise is a finite-size
effect due to a finite sample-to-dimension ratio. The algorithm of
\cite{benaychgeorges2021optimal} is also a rotationally invariant estimator,
i.e.\ one where every eigenvalue is rescaled by a function, though their function
is different from our Eq.~\ref{eq:rescaling}.


Our proof strategy is also different from the one used in the works cited above:
we derive our estimator from a free entropy approach rooted in statistical
physics, which allows us to predict \textit{a priori} the MMSE between the
signal and the denoised sample. This approach was presented for extensive-rank
matrix denoising in \cite{schmidt:tel-03227132,maillard2021perturbative,
barbier2021statistical}. In \cite{maillard2021perturbative} the denoising
problem was solved for square symmetric priors, obtaining the already known
denoiser of \cite{cleaning2017} and predicting its theoretical MMSE. The authors
there study denoising in the context of a high-temperature expansion approach to
matrix factorization. We extend their analysis to non-symmetric rectangular
priors, and we find the symmetric-case denoiser as a special case of our optimal
denoiser. In \cite{barbier2021statistical}, the computation of the denoising
free entropy is presented as a stepping stone towards the computation of minimum
mean-squared error of extensive rank matrix factorization aiming to go beyond
the rotationally invariant priors on $\bS$. The mathematical challenge in
\cite{barbier2021statistical} is linked to finding closed forms for the
corresponding HCIZ integral in the limit of high dimension. The main point of
their work is a closed form expression that presents a conjecture for the free
entropy for matrix denoising and factorization in terms of some spectral
quantities and the HCIZ integral. Evaluating this formula would require to
compute HCIZ in a generic setting, which as far as we know is for the moment out
of reach. If the conjecture of \cite{barbier2021statistical} is correct then our
results can be seen as an explicit evaluation of their formula for the special
case of rotationally invariant priors on $\bS$. Note, however, that even
checking that indeed for rotationally invariant priors the conjecture of
\cite{barbier2021statistical} recovers our results is so far open. 

The asymptotic behaviour of HCIZ integrals --- symmetric and rectangular --- has
been studied extensively in the literature, both in the context of free
probability (\cite{GUIONNET2002461}, \cite{guionnet2021large}) and by the
statistical physics community (\cite{Matytsin,2003HCIZ,collins2020tensor}), yet
explicit forms in special cases were considered only sporadically
\cite{instanton}. To the best of our knowledge, the explicit form given in
\cref{eq.HCIZres} has not been presented elsewhere.

\section{Analytical derivation of the optimal estimator and its
performance}\label{sec.optimal}

As we briefly discussed in the introduction, both the optimal estimator and its
MSE are related to properties of the posterior distribution \cref{eq.posterior3}
and its partition function \cref{eq.partition3}. Moreover, both can be derived
from the knowledge of the free entropy $\Phi_{\bY} \equiv  \log( \caZ_{\bY} ) /
(mp)$. Indeed, one can check that the posterior average, i.e.\ the optimal
estimator, is given by
\begin{equation}\label{eq.estPHI}
    \Sest(\bY)_{i\mu} 
    = \int  d\bS \, \bS_{i\mu}  P(\bS \mid \bY)
    = \bY_{i\mu} +
    \frac{\Delta} {\sqrt{mp} }\frac{1}{\caZ_{\bY}} 
    \frac{\del \caZ_{\bY}}{\del \bY_{i\mu}} 
    = \bY_{i\mu} + \Delta \sqrt{mp}
    \frac{\del \Phi_{\bY} }{\del \bY_{i\mu}} 
    \, ,
\end{equation}
by computing explicitly the derivative $\del_{\bY} \caZ_{\bY}$ starting from
\cref{eq.partition3}. For the MMSE \cref{eq.defMMSE}, by the I-MMSE theorem
\cite{guo2004mutual}, we have
\begin{equation}\label{eq.mmsePHI}
    \MMSE = \Delta + 2 \Delta^2 \del_{\Delta} \EE_{\bY}\left[ \Phi_{\bY} \right]
    \, .
\end{equation}
A sketch of the derivation of \cref{eq.mmsePHI} with our specific normalizations
and notations is given in \cref{app.I-MMSE}. The main focus of the section will
be to compute the free entropy $\Phi_{\bY}$ in the high-dimensional limit.

\subsection{Asymptotics of the free entropy $\Phi_{\bY}$}

To compute the free entropy, we start by performing a change of variable from
$\bS$ to its singular value decomposition $\bS = \bU \bT \bV$ in
\cref{eq.partition3}, with $\bU \in \rotGroup{m}$, $\bV \in \rotGroup{p}$ and
diagonal $\bT \in \bbR_+^{m \times p}$ (recall that without loss of generality
we took $m \le p$). We will denote the singular values as $\bT_{ll} = T_l$ for
$l = 1, \dots, m$. As usual --- see \cite{anderson2010introduction} for example
--- the Jacobian of the change of variable involves the Vandermonde determinant
of the squared singular values $\bT^2$, $\Delta(\bT^2) = \prod_{i<j}^{m} (T_i^2
- T_j^2)$, as well as an additional term involving the product of singular
values\footnote{The change of variable involves also a constant term depending
only on the ratio $p/m = R_1$ that will not contribute to the computation of any
relevant observable. For this reason we can safely avoid computing it
explicitly, and the resulting free entropy will be computed up to
$R_1$-dependent terms.}
\begin{equation}\label{eq:ZY}
    \begin{split}
        \caZ_{\bY} 
        &=
        \Delta^{-\frac{m p}{2}}
        \int \Haar{d \bU}{m}\, \Haar{d \bV}{p} \,d \bT \,
        P_{\rm signal}(\bT) 
        | \Delta(\bT^2) | 
        \prod_{l=1}^m T_l^{p-m}
        \\ &\quad\times
        \exp\left[-\frac{\sqrt{mp}  }{2\Delta} 
        \left(  \sum_{l=1}^{m} T^2_l + \Tr(\bY \bY^T) \right)\right]
        \exp\left[\frac{\sqrt{mp} }{\Delta}
        \Tr\left(  \bU \bT \bV \bY^T \right)\right]
        \\
        &=
        \Delta^{-\frac{m p}{2}} \exp\left[-\frac{\sqrt{mp}  }{2\Delta} 
        \Tr\left(\bY \bY^T\right) \right]
        \int d \bT \,
        P_{\rm signal}(\bT) 
        | \Delta(\bT^2) | 
        \prod_{l=1}^m T_l^{(R_1-1)m}
        \\ &\quad\times
        \exp\left[-\frac{\sqrt{mp}  }{2\Delta} 
        \sum_{l=1}^{m} T^2_l \right]
        \caI_{m}\left(\bT, \bY; \frac{\sqrt R_1}{\Delta} \right)
        \, .
    \end{split}
\end{equation}
where $\Haar{\cdot}{m}$ is the uniform measure on the orthogonal group in
dimension $m$ and where we recognized that the integral over rotation matrices
reduces to the rectangular HCIZ integral defined in \cref{eq.HCIZdef}. Notice
that we used the rotational invariance of the prior to argue that $P_{\rm
signal}(\bS) = P_{\rm signal}(\bT)$. In the non-rotational invariant case
$P_{\rm signal}(\bS)$ would also depend on the singular vectors of $\bS$.

To move forward, we notice that all quantities appearing in the integral depend
on the singular values matrix $\bT$ through its symmetrized empirical singular
value density $\denshatnew{\bT}$, and that they have finite asymptotic limit in
the exponential scale $\exp(m^2)$\footnote{ For the HCIZ, see \cref{sec.HCIZ}.
For the Vandermonde and the product of singular values, we notice that each
product over (which converts into a sum when exponentiating) $i = 1, \dots, m$
contributes to a power $m$ in the exponential scale.}. Then, we change
integration variables from the "$m$-dimensional" empirical density
$\denshatnew{\bT}$ to a generic probability density $\rho$: we argue in
\cref{app.sp} that in the large $m$ limit this change introduces only a term in
the exponential scale $\exp(m)$, which is thus subleading with respect to the
original integrand and can be discarded. Finally, we use Laplace's method to
observe that the integral concentrates in the scale $\exp(m^2)$ onto the value
of the integrand at a deterministic density $\rho^*$; Nishimori identities
\cite{Nishimori_1980} guarantee then that $\rho^* = \denshat{\Sstar}$ (see also
\cite{barbier2021statistical,maillard2021perturbative} for more discussions of
this phenomenon). Thus, by taking the leading asymptotic order of all terms and
disregarding all finite-size corrections and all constant terms depending only
on $R_1$
\begin{equation}\label{eq.phiprehciz}
  \begin{split}
    \Phi_{\bY} 
        &= - \frac{1}{2} \log \Delta
        + \frac{1}{mp} \log P_{\rm signal}[\denshat{\Sstar}] 
        -\frac{1 }{2\Delta\sqrt{R_1} } \Var[\denshat{\Sstar}]
        + \frac{1}{R_1} \Sigma\left[ \denshat{\Sstar}\right]  
        \\&\quad
        + \frac{R_1 - 1}{R_1} \Lambda\left[ \denshat{\Sstar}\right]  
        -\frac{1 }{2\Delta\sqrt{R_1} } \Var[\denshatnew{\bY}]
        +  \frac{1}{2 R_1} 
        I_{R_1}\left[\denshat{\Sstar}, \denshatnew{\bY}; \frac{\sqrt{R_1}}{\Delta}\right]
  \end{split}
\end{equation}
where $\Var[\sigma] = \int dx\, \sigma(x) x^2$, $\Sigma[\sigma] = \dashint dx \,
dy \, \sigma(x) \sigma(y)  \log |x - y|$ and $\Lambda[\sigma] = \dashint dx \,
\sigma(x) \log |x|$, and $\denshat{\Sstar}$ is the asymptotic symmetrized
singular value densities of $\Sstar$. $\denshatnew{\bY}$ instead is the
symmetrized empirical density of singular values of the fixed instance of the
observation $\bY$. All dashed integrals are regularized as detailed in
\cref{app.regularization}.

In \cref{eq.phiprehciz} we took the high-dimensional limit of the prior term
$(mp)^{-1} \log P_{\rm signal}(\bT)$ in a rather uncontrolled way. Treating the
limit rigorously is delicate, see for example the discussions in \cite[Section
II.C]{barbier2021statistical}, but brings no surprises for non-pathological
priors --- for example, the factorized priors in the symmetric version of our
problem \cite{maillard2021perturbative}. As mentioned at the start of the
section, all quantities we are interested in depend on derivatives of the free
entropy with respect to $\Delta$ or to $\bY$, and the prior term brings no
contribution at all in these cases. For this reason, we do not treat it in
detail. 

We will show in \cref{sec.HCIZ} that at leading order
\begin{equation}
    \begin{split}
    I_{R_1} \left[\denshat{\Sstar}, \denshatnew{\bY}; \frac{\sqrt{R_1}}{\Delta}\right]
    &= C_{R_1} + R_1 \log\Delta
    + \frac{\sqrt{R_1}}{\Delta} \Var[\denshat{\Sstar}] 
    + \frac{\sqrt{R_1}}{\Delta} \Var[\denshatnew{\bY}] 
    \\&\quad
    - 2 (R_1 - 1) \Lambda[\denshatnew{\bY}] 
    - 2 \Sigma[\denshatnew{\bY}]
    \end{split}  
\end{equation}
for some constant $C_{R_1}$ depending only on $R_1$. The asymptotic free entropy
thus equals 
\begin{equation}\label{eq.freeentropy}
    \begin{split}        
        \Phi_{\bY} 
        &= \text{const}(R_1) 
        + \frac{1}{mp} \log P_{\rm signal}[\denshat{\Sstar}] 
        + \frac{1}{R_1} \Sigma\left[ \denshat{\Sstar}\right]  
        + \frac{R_1 - 1}{R_1} \Lambda\left[ \denshat{\Sstar}\right] 
        \\&\quad
        - \frac{1}{R_1} \Sigma\left[ \denshatnew{\bY}\right]      
        - \frac{R_1 - 1}{R_1} \Lambda[\denshatnew{\bY}]
        \, ,
    \end{split}
\end{equation}
where $\text{const}(R_1)$ gathers all constants depending on $R_1$ that we did
not trat explicitly, namely the constant in the HCIZ asymptotics and the
constant in the intial SVD change of variable. Notice that only the last two
terms depend either on $\bY$ or on $\Delta$ (through the spectral properties of
$\bY$).

\subsection{Explicit form of the MMSE and the optimal estimator}

The MMSE can be computed directly by combining \cref{eq.mmsePHI} with
\cref{eq.freeentropy}, obtaining
\begin{equation}
  \begin{split}
      \MMSE
      = \Delta - 2 \Delta^2 \frac{\del}{\del \Delta} \EE_{\bY}
      \left[  \frac{1}{R_1}  \Sigma[ \denshatnew{\bY}] 
      + \frac{R_1 - 1}{R_1}  \Lambda[ \denshatnew{\bY}] \right]
  \, .
  \end{split}
\end{equation}
Thanks to the concentration of spectral densities in the high-dimensional limit,
the average over $\bY$ can be performed directly by substituting the
\textit{empirical} symmetrized singular value density $\denshatnew{\bY}$ of the
specific instance of the observation $\bY$ with the \textit{asymptotic} singular
value density $\denshat{\bY}$ of the observation, which is a deterministic
quantity: it depends only on the statistical properties of $\bY$, and not on its
specific value. Thus, we obtain \cref{res.mmse}.

To compute the explicit form of the optimal estimator, we start from
\cref{eq.estPHI} and notice that the free entropy depends only on spectral
properties of the actual instance of the observation $\bY$. Thus, the derivative
of the free entropy w.r.t. one component of $\bY$ can be decomposed on the
eigenvalues of $\bY$ as follows
\begin{equation}
  \Sest(\bY)_{i\mu} 
  = \bY_{i\mu} + \Delta \sqrt{mp}
  \frac{\del \Phi_{\bY} }{\del \bY_{i\mu}} 
  = \bY_{i\mu} + \Delta \sqrt{mp}
    \sum_{l=1}^{m} \frac{\del \Phi_{\bY} }{\del y_l} \frac{\del y_l}{\del \bY_{i\mu}} 
  \, ,
\end{equation}
where in the last passage we called $\{y_i\}_{i=1}^{m}$ the singular values of
$\bY$.

To compute the derivative, we use a variant of the Hellman-Feynman theorem
\cite{Cohen-Tannoudji:101367}. The Hellman-Feynman theorem considers a symmetric
matrix depending on a parameter, and relates the derivative of an eigenvalue of
the matrix with respect to that parameter with the derivative of the matrix
itself. In \cref{app.HF} we show that the Hellman-Feynman theorem can be adapted
to the case of rectangular matrices and singular values, and in particular we
show that $\del_\lambda y = u^T \cdot \del_\lambda {\bM} \cdot v$, if $\bM$ is a
matrix, $y$ one of its singular values, $u$ and $v$ the corresponding left and
right singular vectors, and all these quantities depend on a parameter
$\lambda$. 

To compute $\del y_{l} / \del \bY_{i_0 \mu_0} $ we need to perturb the original
matrix $\bY$ as $\bY(\lambda)_{\mu i} = \bY_{i\mu} + \lambda \delta_{i i_0}
\delta_{\mu \mu_0}$ so that (here $u^l$ and $v^l$ are the left and right
singular values of $\bY$ corresponding to the $l$-th singular vector)
\begin{equation}
    \begin{split}
        \frac{\del y_{l} }{\del \bY_{i_0 \mu_0}} 
        &= \frac{\del y_{l} }{\del \lambda} 
        =
         (u^l)^T \cdot \del_\lambda \bY(\lambda) \cdot v^l
        =
        \sum_{i, \mu = 1}^{m, p}  u_i^l \delta_{i i_0} \delta_{\mu \mu_0} v^l_\mu
        = u_{i_0}^l v_{\mu_0}^l  \, .
    \end{split}
\end{equation}
For clarity, $u_i^l$ is the i-th component of the left singular vector
corresponding to the $l$-th singular value, and similarly for $v_\mu^l$. Thus
\begin{equation}
    \left\langle \bS_{i \mu} \right\rangle
    = \bY_{i\mu} + \Delta \sqrt{mp}
    \sum_{l=1  }^{m}
    \frac{\del \Phi_{\bY} }{\del y_{l}} 
    u^l_{i} v^l_{\mu} 
    = 
    \sum_{l=1  }^{m}
    \left[
        y_{l} +
        \Delta \sqrt{mp} \frac{\del \Phi_{\bY} }{\del y_{l}} 
    \right]
    u^l_{i} v^l_{\mu} 
    = 
    \sum_{l=1  }^{m}
    \xi( y_{l} )   
    u^l_{i} v^l_{\mu} 
    \, ,
\end{equation}
where we used that $\bY_{i\mu} = \sum_{l=1}^{m} y_l u^l_i v^l_\mu$ by definition
of SVD and we introduced the spectral denoising function $\xi(y)$. This proves
the first claim of \cref{res.est}, i.e.\ that the optimal denoiser is diagonal
in the bases of left and right singular vectors of the observation $\bY$.

To obtain an explicit form for the spectral denoising function $\xi$, we need to
compute $\del \Phi_{\bY} /\del y_{l} $. For this, we consider the part of the
free entropy depending on $\bY$, call it $\Psi_{\bY}$, in the discrete setting
(all following equalities are intended at leading order for large $m$), i.e.
\begin{equation}
    \begin{split}   
        \Psi_{\bY} &=          
        - \frac{R_1 - 1}{R_1} \frac{1}{m} \sum_{l : y_l \neq 0} \log|y_l|
        - \frac{1}{R_1} \frac{1}{m^2} \sum_{l\neq l'}  \log |y_l - y_{l'}|
        \, ,
    \end{split}
\end{equation}
so that
\begin{equation}
    \begin{split}   
        \sqrt{mp} \frac{\del \Psi_{\bY} }{\del y_{l}}   
        &=
        - \frac{R_1 - 1}{y_l\sqrt {R_1}} 
         - \frac{2}{\sqrt{R_1}} \frac{1}{m} \sum_{k} \frac{1}{y_l - y_{k}} 
       = 
        - \frac{1}{\sqrt {R_1}}  \frac{R_1 - 1}{y_l}
        - \frac{2}{\sqrt {R_1}} \dashint dx \,  \frac{\denshatnew{\bY}(x)}{y_l - x}
        \, ,
    \end{split}
\end{equation}
and the spectral denoising function is finally given by 
\begin{equation}
  \xi(y) =y - \frac{2 \Delta}{\sqrt{R_1}} \left[ \frac{R_1 -1}{2y} + \dashint d\zeta\, 
  \frac{\denshatnew{\bY}(\zeta)}{y-\zeta} \right] \, .
\end{equation}
We thus recover the second part of \cref{res.est}, i.e.\ the expression for the
denoising function $\xi$, by invoking once again concentration of spectral
densities, so that $\denshatnew{\bY}$ can be safely substituted by its
asymptotic deterministic counterpart $\denshat{\bY}$ in the high-dimensional
limit.

\section{Specific priors}\label{sec.priors}

In this section we provide all the ingredients needed to specialize
\cref{res.est} and \cref{res.mmse} to specific choices of rotationally-invariant
priors. We then analyze in detail the case of the Gaussian factorized matrix
prior defined in \cref{eq.defWishart}.

\subsection{General remarks}\label{sec.general}

The two main ingredients needed to make our results explicit for a specific form
of the prior are the symmetrized singular value density of the observation
$\denshat{\bY}$ and its Hilbert transform $\mathcal{H}[\denshat{\bY}](y) =
\frac{1}{\pi}\dashint d\zeta\, \denshat{\bY}(\zeta) / (y-\zeta)$. Now we show
how to compute them analytically. 

The first ingredient needed is the so-called \textit{Stieltjes transform} of the
symmetrized singular value density $\denshat{\bA}(x)$, i.e.
\begin{equation}
    g_{\bA}(z) = \int dx \frac{\denshat{\bA}(x)}{z-x}
    \quad \text{for} \quad z \in \bbC \setminus \supp{\denshat{\bA}}
    \, ,
\end{equation}
where $\supp{\denshat{\bA}}$ denotes the support $\denshat{\bA}$. Indeed, one
can show that (see for example \cite{potters_bouchaud_2020})
\begin{equation}
    \denshat{\bA}(x) = \frac{1}{\pi} \lim_{\epsilon \to 0^+} \Im g_{\bA}(x - i \epsilon) 
     \quad \text{and} \quad
    \dashint dx \,  \frac{\denshat{\bA}(x)}{y-x} = 
    \lim_{\epsilon \to 0^+} \Re g_{\bA}(x - i \epsilon)
\end{equation}
where the second equality follows from Kramers–Kronig relations
(\cite{Jackson:100964}). Thus, from the knowledge of $g_{\bA}$ one can recover
both the symmetrized singular value density and the corresponding Hilbert
transform, obtaining all the ingredients needed to make the optimal estimator
Eq.~\eqref{eq.res1} and the MMSE Eq.~\eqref{eq.res2} explicit for a given prior.
Notice that, while in the following we will focus on cases in which $g_{\bA}$
can be computed analytically, our algorithm can be in principle applied to any
rotationally-invariant ensemble of random matrices $\bA$ by sampling a large
matrix and computing its singular values to estimate $g_{\bA}$ as in
\cite{ledoit2009eigenvectors}.

Thus, the questions shifts to how to compute the Stieltjes transform of
$\denshat{\bA}$, which in turn entails two sub-problems: determining the
Stieltjes transform of the prior $g_{\Sstar}(z)$, and then adding the effect of
the noise.

In order to compute the Stieltjes transform of the prior\footnote{ $\bA =
  \Sstar$ in the following as the reasoning holds for generic matrices $\bA$. },
  we consider the matrix $\bA \bA^T$. Indeed, if $\bA$ has SVD $\bA = \bU \bB
  \bV$ with diagonal $\bB$, then $\bA \bA^T = \bU \bB \bB^T \bU^T$, from which
  we see that there is a one-to-one correspondence between singular values of
  $\bA$ and eigenvalues of $\bA \bA^T$. Namely the former are the positive
  square roots of the latter (notice that $\bA\bA^T$ is symmetric and positive
  semi-definite)\footnote{One needs to be careful here: we used that $m \leq p$
  and $\bA \in \bbR^{m \times p}$, so that $\bA \bA^T$ has as much eigenvalues
  as the number of singular values of $\bA$. In general, one needs to choose
  between $\bA \bA^T$ and $\bA^T \bA$ the one with lower dimensionality in order
  not to add spurious null singular values}. This eigenvalues-singular values
  relations has a consequence on the corresponding Stieltjes transforms. Indeed
\begin{equation}\label{eq.relG}
    g_{\bA}(z) 
    = \int dx \, \frac{\denshat{\bA}(x)}{z - x}
    = \int_0^\infty dx \,  \frac{2z\,\dens{\bA}(x)}{z^2 - x^2}
    = \int_0^\infty d\lambda \,  \frac{z \, \dens{\bA\bA^T}(\lambda)}{z^2 - \lambda}
    = z g_{\bA\bA^T}(z^2) 
\end{equation}
where $\dens{\bA\bA^T}$ is the usual spectral density of $\bA\bA^T$. Thus,
\cref{eq.relG} links the Stieltjes transform of the singular value density of a
rectangular matrix, which is in principle a very generic and non-trivial
quantity to compute, to the Stieltjes transform of the eigenvalue density of a
symmetric square matrix, for which many explicit analytical and numerical
results already exist in the literature \cite{potters_bouchaud_2020}.

Notice that the procedure just described is of no help in order to add the
noise: in fact, $(\bA + \bZ)(\bA + \bZ)^T$ is a sum of products of matrices that
are not independent (free) between each other. Thus, usual free probability
techniques based on the $\caR$ or $\caS$ transforms and free convolution of
symmetric matrices would fail to treat this problem.

In order to add the effect of the noise, one needs \textit{rectangular free
probability} techniques. We do not enter into the details here: we refer the
interested reader to \cite{Benaych-Georges2009}. The underlying idea is not
different from what happens in the more conventional symmetric case: there
exists a functional transform of the Stieltjes transform, the rectangular $\caR$
transform, that linearizes the sum of random matrices, i.e. $\caR[g_{\bA}] +
\caR[g_{\bB}] = \caR[g_{\bA + \bB}]$ whenever the random matrices $\bA$ and
$\bB$ are mutually free. Thus, one can access the Stieltjes transform of the sum
by computing two direct $\caR$ transforms, and one inverse $\caR$ transform. In
practice, the computation of the rectangular $\caR$ transform entails the
numerical estimation of functional inverses as soon as one departs from the
simplest case of matrices with i.i.d.\ Gaussian entries. For this reason, in the
special case of factorized priors \cref{eq.defWishart}, we will not pursue this
very general strategy. Instead, in \cref{sec.factorized} we will adapt the
results of \cite{NIPS2017_0f3d014e, Louart}, which are easier to implement
numerically.

As a final remark, notice that the Stieltjes transform scales under a rescaling
of $\bA$ as $g_{c\bA}(z) = g_{\bA}\left(z/c\right) / c$. This will be useful to
deal with the many constant terms appearing in our computations.

\subsection{Symmetric priors}

The case of symmetric priors has already been treated originally in
\cite{RIEoriginal}, and more recently in \cite{maillard2021perturbative} with
techniques akin to ours. In particular, with a symmetric prior one can repeat
our analysis using eigenvalue densities instead of singular value densities. For
positive semi-definite symmetric priors, it is immediate to see that our results
reduce to those of \cite{RIEoriginal,maillard2021perturbative}, as singular
values coincide with eigenvalues in this case. For generic symmetric priors,
singular values are the absolute values of eigenvalues, and thus the singular
value density is simply given by the symmetrized spectral density.

\subsection{The case of Gaussian factorized prior}\label{sec.factorized}

The prior we would like to focus our attention on is given by the model of
Gaussian extensive-rank factorized signals defined in \cref{eq.defWishart}. As
argued in \cref{sec.general}, we need a way to compute the Stieltjes transform
of the observation $\bY$ for this specific prior in order to have access to
$\denshat{\bY}$ --- to numerically compute the integrals in the MMSE formula
\cref{eq.res2} --- and its Hilbert transform --- to be able to perform actual
denoising on given instances of $\bY$ using \cref{eq.res1}.

To this end, we generalize the results given in \cite{NIPS2017_0f3d014e}. There,
the authors study the spectrum of $\bC = f(\bF \bX) f(\bF \bX)^T$, where $f$ is
a component-wise non-linearity and $\bF, \bX$ matrices with i.i.d.\ Gaussian
entries. They show that the Stieltjes transform of the spectral density of $\bC$
satisfies a degree-four algebraic equation depending only on the aspect-ratio
$R_1$, the rank parameter $R_2$ and two Gaussian moments of $f$, $\eta = \int Dz
f(z)^2$ and $\zeta = \big[\int Dz f'(z)\big]^2$, where $Dz$ denotes integration
over a standard Gaussian random variable.

Recall that, thanks to \cref{eq.relG}, knowing the Stieltjes transform of the
spectral density of $\bC$ is equivalent to knowing that of the singular value
density of $f(\bF \bX)$. Thus, we could compute $g_{\bY}$ by choosing a noisy
function $f(x) = x + z$, where $z$ is a Gaussian random variable (actually, one
for each component of the matrix, all i.i.d.). In order to do that, we just need
to show that the results of \cite{NIPS2017_0f3d014e} extend to non-deterministic
non-linearities --- only the deterministic case is considered therein. This
happens to be the case: we provide the details of the extension in
\cref{app.proofPennington}. The only effect of the noise of the non-linearity is
a redefinition of the Gaussian moments of $f$, and in particular $\eta_{\rm
noisy} = \EE_f \int Dz f(z)^2$ and $\zeta_{\rm noisy} = \EE_f\left[\int Dz
f'(z)\right]^2$, where $\EE_f$ denotes averaging over the noise induced by $f$.

Thus, we can safely use the results from \cite{NIPS2017_0f3d014e} to compute the
Stieltjes transform of the singular value density of $\bY$ for Gaussian
factorized priors. There is a non-trivial mismatch between our normalizations
and those of \cite{NIPS2017_0f3d014e}: to fall-back directly onto their results,
we need to choose $f(x) = \sqrt[4]{R_1} (x + \sqrt{\Delta} z)$, and set their
parameters to $\sigma_x = \sigma_w = 1$, $\phi = R_2/R_1$ and $\psi = R_2$.

With these choices, the Gaussian moments of $f$ are given by $\eta_{\rm noisy} =
(1+\Delta) \sqrt{R_1}$ and $\zeta_{\rm noisy} = \sqrt{R_1}$. The algebraic
equation for $g_{\bY}(z)$ is given by \cite[SM, Eq. S42-S43]{NIPS2017_0f3d014e}
and is reported in \cref{app.proofPennington}.

\section{Asymptotics of the HCIZ integral}\label{sec.HCIZ}

In this section we consider the asymptotics of the rectangular HCIZ presented in
\cite{guionnet2021large} and adapt them to the problem of denoising.

We follow the notations of the paper. In particular, their parameters $\alpha$
and $\beta$ in our case equal $\alpha = R_1 -1$ and $\beta = 1$. The HCIZ
integral was defined in \cref{eq.HCIZdef},
and its asymptotic value is defined as
\begin{equation}
    I_{\alpha}[\denshat{\bA}, \denshat{\bB}; \lambda] =
    \lim_{m \to \infty, p = (1+\alpha) m} 
    \frac{2}{m^2} \log \caI_{m}(\bA, \bB ; \lambda)
    \, ,
\end{equation}
where $\denshat{\bA, \bB}$ are the asymptotic symmetrized singular value
densities of $\bA$ and $\bB$. Following \cite[Theorem 1.1]{guionnet2021large},
the asymptotic of the HCIZ integral with parameter $\tau = 1$ equals 
\begin{equation}\label{eq.action}
    \begin{split}
        I_{\alpha} \left[\denshat{\bA}, \denshat{\bB}; \tau = 1\right]
        &= 
        C_{\alpha} 
        + \Var[\denshat{\bA}] - \alpha \Lambda[\denshat{\bA}] - \Sigma[\denshat{\bA}]
        + \Var[\denshat{\bB}] - \alpha \Lambda[\denshat{\bB}] - \Sigma[\denshat{\bB}]
        \\
        &\quad
        -\inf_{\{\hrho_t\}_{0\leq t \leq 1}} \left\{ 
            \int_0^1 ds \int dx \hrho_s (x) \left[
                v_s^2(x) + \frac{\pi^2}{3} \hrho_s^2(x) + \frac{\alpha^2}{4 x^2}
            \right] 
        \right\}  
    \end{split}  
\end{equation}
where $\Var[\sigma] = \int dx \, \sigma(x) x^2$, $\Lambda[\sigma] = \int dx \,
\sigma(x) \log|x|$, $\Sigma[\sigma] = \dashint dx dy \, \sigma(x) \sigma(y)
\log|x-y|$, $C_\alpha$ is an unspecified constant depending only on $\alpha$,
the infimum is taken over continuous symmetric density valued processes
$(\hrho_t(x))_{0\leq t \leq 1}$ such that $\hrho_0(x) = \denshat{\bA}$ and
$\hrho_1(x) = \denshat{\bB}$, and where $v_t(x)$ is a solution to the continuity
equation $\del_t \hrho_t (x) + \del_x \left( \hrho_t (x) v_t(x) \right) = 0 $.
We refer the reader to the original work for precise definitions and assumptions
over all quantities.

The non-trivial part of Eq.~\eqref{eq.action}, i.e.\ the optimization problem,
is a mass transport problem whose solution interpolates between the singular
value densities of the two matrices $\bA$ and $\bB$. This transport problem can
be understood as the evolution of a hydrodynamical system, and it can be shown
\cite[Eq. 4.13]{menon2017complex} that the resulting density profile
$\hrho_t(x)$ (i.e. the process at which the infimum is reached) is the
symmetrized singular value density of the Dyson Brownian bridge between $\bA$
and $\bB$, i.e.\ the symmetrized singular value density of the matrix
\begin{equation}\label{eq.bridge}
    \bX(t) = (1-t) \bA + t \bB + \sqrt{t(1-t)} \bW
\end{equation}
where $\bW$ is a matrix of i.i.d.\ Gaussian variables with mean zero and
variance $1/m$. It can be shown \cite[section 4]{guionnet2021large} that  
the velocity field $v_t(x)$ satisfies
\begin{equation}\label{eq.vel}
    v_t(x) = \pi \mathcal{H}[\hrho_t](x) + \frac{\alpha}{2 x} + D(x, t)
\, , \qquad
 \mathcal{H}[\hrho](x) = \frac{1}{\pi} \dashint dy\, \frac{\hrho(y)}{x-y},
\end{equation}
where $\mathcal{H}[\cdot]$ is the Hilbert transform, $\dashint$ the Cauchy
principal value integral, and $D(x,t)$ is a \textit{drift} term that ensures
that the Brownian motion ends precisely at $\bB$ when $t=1$.

Our aim is to make \cref{eq.action} explicit in the specific case $\bB - \bA =
\kappa \bW$, where $\bW$ is a matrix of i.i.d.\ Gaussian variables with mean
zero and variance $1/m$ and $\kappa$ a positive constant, and for this specific
case to extend \cref{eq.action} to $\tau \neq 1$.
One could in principle absorb $\tau$ in the
definition of the matrix $\bA$, or $\bB$, or both. We will see shortly that if
one wants to make more the asymptotic form of the HCIZ integral explicit in the
special case of $\bB = \bA$ + Gaussian noise --- relevant for the denoising
problem --- one needs to be careful. In particular, the correct way to absorb
$\tau$ is to split it evenly between $\bA$ and $\bB$, by defining $\bA' =
\sqrt{\tau} \bA$ and $\bB' = \sqrt{\tau} \bB$. In this way, the fact that the
difference between $\bB'$ and $\bA'$ is a matrix with Gaussian i.i.d. entries
--- which, as we will see shortly, is a key property --- gets preserved. Thus, $
I_{\alpha} [\denshat{\bA}, \denshat{\bA + \kappa \bW}; \tau ] = I_{\alpha}
[\denshat{\bA'}, \denshat{ \bA' + \sqrt{\tau} \kappa \bW}; 1 ] $.
  
Now we notice that in the particular case of $\bB' = \bA' + \sqrt{\tau} \kappa
\bW$, \cref{eq.bridge} simplifies to a Dyson Brownian motion starting at $\bA$
\begin{equation}
    \begin{split}
        \bX(t) 
        &= (1-t) \bA' + t (\bA' + \sqrt{\tau} \kappa \bW_1) + \sqrt{t(1-t)} \bW_2 = \bA' + \sqrt{t} \bW 
    \end{split},
\end{equation}
where we summed the two independent (more precisely, free) realizations of the
Gaussian matrix $\bW_1$ and $\bW_2$ by substituting them with a third
independent Gaussian matrix $\bW$ and by summing the original variances. We also
used the fact that $\sqrt{\tau} \kappa = 1$ in the denoising problem; we will
discuss the $\sqrt{\tau} \kappa \neq 1$ briefly at the end of this section.
Moreover, no drift is needed to impose that $\bX(1) = \bB'$, so that in
\cref{eq.vel} $D(x, t) \equiv 0$ and
\begin{equation}\label{eq.vel2}
  v_t(x) = \pi \mathcal{H}[\hrho_t](x) + \frac{\alpha}{2 x} \, .
\end{equation}

The explicit form \cref{eq.vel2} is crucial to make explicit
Eq.~\eqref{eq.action}. Indeed, one can show that if \cref{eq.vel2} holds, then
\begin{equation}\label{eq.Gt}
    \begin{split}
        G(t) &=
        \Sigma[ \hrho_t ] + \alpha \Lambda[ \hrho_t ]
        - \int_0^t ds \int dx \hrho_s (x) \left[
            v_s^2(x) + \frac{\pi^2}{3} \hrho_s^2(x) + \frac{\alpha^2}{4 x^2}
        \right] 
    \end{split}
\end{equation}
is a constant function of $t$, i.e.\ $\del_t G(t) = 0$, see \cref{app.hciz}. The
fact that $G(0) = G(1)$ allows to explicitly write the dynamical portion of
Eq.~\eqref{eq.action} as
\begin{equation}
        \int_0^1 ds \int dx \hrho_s (x) \left[
        v_s^2(x) + \frac{\pi^2}{3} \hrho_s^2(x) + \frac{\alpha^2}{4 x^2}
        \right] 
        = \Sigma[ \denshat{\bA'} ] + \alpha \Lambda[ \denshat{\bA'} ] 
        - \Sigma[ \denshat{\bB'} ] - \alpha \Lambda[ \denshat{\bB'} ]
\end{equation}
in all cases in which $\bB' = \bA' + \bW$ where again $\bW$ is a matrix of
i.i.d.\ Gaussian variables with mean zero and variance $1/m$.

To sum it up and get back to the case of denoising, we have, calling $\tau =
\sqrt{R_1} / \Delta$, 
\begin{equation}
    \begin{split}
    I_{R_1} & \left[\denshat{\bS}, \denshat{\bY}; \tau = \frac{\sqrt{R_1}}{\Delta}\right]
    = 
    I_{R_1} \left[\denshat{\sqrt{\tau} \bS}, 
    \denshat{\sqrt{\tau} \bS + \bW}; 1 \right]
    \\
    &= 
    C_{R_1} + R_1 \log\Delta
    + \frac{\sqrt{R_1}}{\Delta} \Var[\denshat{\bS}] 
    + \frac{\sqrt{R_1}}{\Delta} \Var[\denshat{\bY}] 
    - 2 (R_1 - 1) \Lambda[\denshat{\bY}] 
    - 2 \Sigma[\denshat{\bY}] \, ,
    \end{split}
\end{equation}
where $C_{R_1}$ hides all constants that depend exclusively on $R_1$, and we
used that $\denshat{\sqrt{\tau}\bM}(x) = \denshat{\bM}\left(x /
\sqrt{\tau}\right) / \sqrt{\tau}$.

As a final remark, in the case in which $\sqrt{\tau} \kappa \neq 1$, one can
replicate exactly this argument by rescaling the time domain so that $\bB' =
\bX(t = \sqrt{\tau} \kappa)$. This does not change anything, apart from possibly
modifying the unspecified constant $C_{R_1}$ into a constant $C(R_1, \sqrt{\tau}
\kappa)$. This concludes our derivation of \cref{res.hciz}.


\section*{Acknowledgements}
 We thank Marc Mézard and Francesco Camilli for discussions about matrix factorization and HCIZ integrals. We thank Jean-Philippe Bouchaud and Alice Guionnet for discussions and pointers to related literature. We acknowledge funding from the ERC under the European Union’s Horizon 2020 Research and Innovation Programme Grant Agreement 714608-SMiLe.

\bibliography{bibliography.bib}

\begin{thebibliography}{35}
\providecommand{\natexlab}[1]{#1}
\providecommand{\url}[1]{\texttt{#1}}
\expandafter\ifx\csname urlstyle\endcsname\relax
  \providecommand{\doi}[1]{doi: #1}\else
  \providecommand{\doi}{doi: \begingroup \urlstyle{rm}\Url}\fi

\bibitem[Bun et~al.(2017)Bun, Bouchaud, and Potters]{cleaning2017}
Jo\"el Bun, Jean-Philippe Bouchaud, and Marc Potters.
\newblock Cleaning large correlation matrices: Tools from random matrix theory.
\newblock \emph{Physics Reports}, 666:\penalty0 1–109, 2017.

\bibitem[Johnstone and Lu(2009)]{sparsePCA}
Iain~M. Johnstone and Arthur~Yu Lu.
\newblock On consistency and sparsity for principal components analysis in high
  dimensions.
\newblock \emph{Journal of the American Statistical Association}, 104\penalty0
  (486):\penalty0 682--693, 2009.

\bibitem[Mairal et~al.(2009)Mairal, Bach, Ponce, and Sapiro]{online_dictionary}
Julien Mairal, Francis Bach, Jean Ponce, and Guillermo Sapiro.
\newblock Online dictionary learning for sparse coding.
\newblock In \emph{Proceedings of the 26th Annual International Conference on
  Machine Learning}, ICML '09, page 689–696, 2009.

\bibitem[Lesieur et~al.(2017)Lesieur, Krzakala, and Zdeborová]{2017Thibault}
Thibault Lesieur, Florent Krzakala, and Lenka Zdeborová.
\newblock Constrained low-rank matrix estimation: phase transitions,
  approximate message passing and applications.
\newblock \emph{Journal of Statistical Mechanics: Theory and Experiment},
  2017\penalty0 (7):\penalty0 073403, 2017.

\bibitem[Miolane(2018)]{miolane2018fundamental}
Léo Miolane.
\newblock Fundamental limits of low-rank matrix estimation: the non-symmetric
  case.
\newblock \emph{Preprint arXiv:1702.00473}, 2018.

\bibitem[Kabashima et~al.(2016)Kabashima, Krzakala, Mezard, Sakata, and
  Zdeborova]{kabashima}
Yoshiyuki Kabashima, Florent Krzakala, Marc Mezard, Ayaka Sakata, and Lenka
  Zdeborova.
\newblock Phase transitions and sample complexity in bayes-optimal matrix
  factorization.
\newblock \emph{IEEE Transactions on Information Theory}, 62\penalty0
  (7):\penalty0 4228–4265, 2016.

\bibitem[Schmidt(2018)]{schmidt:tel-03227132}
Hinnerk~Christian Schmidt.
\newblock \emph{{Statistical Physics of Sparse and Dense Models in Optimization
  and Inference}}.
\newblock PhD thesis, {Universit{\'e} Paris Saclay}, 2018.

\bibitem[Barbier and Macris(2021)]{barbier2021statistical}
Jean Barbier and Nicolas Macris.
\newblock Statistical limits of dictionary learning: random matrix theory and
  the spectral replica method.
\newblock \emph{Preprint arXiv:2109.06610}, 2021.

\bibitem[Maillard et~al.(2021)Maillard, Krzakala, Mézard, and
  Zdeborová]{maillard2021perturbative}
Antoine Maillard, Florent Krzakala, Marc Mézard, and Lenka Zdeborová.
\newblock Perturbative construction of mean-field equations in extensive-rank
  matrix factorization and denoising.
\newblock \emph{Preprint arXiv:2110.08775}, 2021.

\bibitem[Harish-Chandra(1957)]{HC}
Harish-Chandra.
\newblock Differential {O}perators on a {S}emisimple {L}ie {A}lgebra.
\newblock \emph{American Journal of Mathematics}, 79\penalty0 (1):\penalty0
  87--120, 1957.

\bibitem[Itzykson and Zuber(1980)]{IZ}
C.~Itzykson and J.‐B. Zuber.
\newblock The planar approximation. {II}.
\newblock \emph{Journal of Mathematical Physics}, 21\penalty0 (3):\penalty0
  411--421, 1980.

\bibitem[Guionnet and Huang(2021)]{guionnet2021large}
Alice Guionnet and Jiaoyang Huang.
\newblock Large deviations asymptotics of rectangular spherical integral.
\newblock \emph{Preprint arXiv:2106.07146}, 2021.

\bibitem[Cover(1999)]{cover1999elements}
Thomas~M Cover.
\newblock \emph{Elements of information theory}.
\newblock John Wiley \& Sons, 1999.

\bibitem[Guo et~al.(2005)Guo, Shamai, and Verdu]{guo2004mutual}
Dongning Guo, S.~Shamai, and S.~Verdu.
\newblock Mutual information and minimum mean-square error in {G}aussian
  channels.
\newblock \emph{IEEE Transactions on Information Theory}, 51\penalty0
  (4):\penalty0 1261--1282, 2005.

\bibitem[Nishimori(1980)]{Nishimori_1980}
H~Nishimori.
\newblock Exact results and critical properties of the {I}sing model with
  competing interactions.
\newblock \emph{Journal of Physics C: Solid State Physics}, 13\penalty0
  (21):\penalty0 4071--4076, 1980.

\bibitem[Matytsin(1994)]{Matytsin}
A.~Matytsin.
\newblock On the large-{N} limit of the {I}tzykson-{Z}uber integral.
\newblock \emph{Nuclear Physics B}, 411\penalty0 (2–3):\penalty0 805–820,
  1994.

\bibitem[Baik et~al.(2005)Baik, Arous, and Péché]{baik2004phase}
Jinho Baik, Gérard~Ben Arous, and Sandrine Péché.
\newblock {Phase transition of the largest eigenvalue for nonnull complex
  sample covariance matrices}.
\newblock \emph{The Annals of Probability}, 33\penalty0 (5):\penalty0 1643 --
  1697, 2005.

\bibitem[Ledoit and Wolf(2012)]{2012Ledoit}
Olivier Ledoit and Michael Wolf.
\newblock Nonlinear shrinkage estimation of large-dimensional covariance
  matrices.
\newblock \emph{The Annals of Statistics}, 40\penalty0 (2), 2012.

\bibitem[Bun et~al.(2016)Bun, Allez, Bouchaud, and Potters]{RIEoriginal}
Jo\"el Bun, Romain Allez, Jean-Philippe Bouchaud, and Marc Potters.
\newblock Rotational invariant estimator for general noisy matrices.
\newblock \emph{IEEE Transactions on Information Theory}, 62\penalty0
  (12):\penalty0 7475--7490, 2016.

\bibitem[Benaych-Georges et~al.(2021)Benaych-Georges, Bouchaud, and
  Potters]{benaychgeorges2021optimal}
Florent Benaych-Georges, Jean-Philippe Bouchaud, and Marc Potters.
\newblock Optimal cleaning for singular values of cross-covariance matrices.
\newblock \emph{Preprint arXiv:1901.05543}, 2021.

\bibitem[Guionnet and Zeitouni(2002)]{GUIONNET2002461}
Alice Guionnet and Ofer Zeitouni.
\newblock Large deviations asymptotics for spherical integrals.
\newblock \emph{Journal of Functional Analysis}, 188\penalty0 (2):\penalty0
  461--515, 2002.

\bibitem[Zinn-Justin and Zuber(2003)]{2003HCIZ}
P~Zinn-Justin and J-B Zuber.
\newblock On some integrals over the {U(N)} unitary group and their large {N}
  limit.
\newblock \emph{Journal of Physics A: Mathematical and General}, 36\penalty0
  (12):\penalty0 3173–3193, 2003.

\bibitem[Collins et~al.(2020)Collins, Gurau, and Lionni]{collins2020tensor}
Benoît Collins, Razvan Gurau, and Luca Lionni.
\newblock The tensor {H}arish-{C}handra-{I}tzykson-{Z}uber integral {I}:
  {W}eingarten calculus and a generalization of monotone {H}urwitz numbers.
\newblock \emph{Preprint arXiv:2010.13661}, 2020.

\bibitem[Bun et~al.(2014)Bun, Bouchaud, and Majumdar]{instanton}
J.~Bun, J.P. Bouchaud, and M.~Majumdar, S.N.~Potters.
\newblock Instanton approach to large n harish-chandra-itzykson-zuber
  integrals.
\newblock \emph{Physical Review Letters}, 113\penalty0 (7), 2014.

\bibitem[Anderson et~al.(2010)Anderson, Guionnet, and
  Zeitouni]{anderson2010introduction}
Greg~W Anderson, Alice Guionnet, and Ofer Zeitouni.
\newblock \emph{An introduction to random matrices}.
\newblock Number 118. Cambridge university press, 2010.

\bibitem[Cohen-Tannoudji et~al.(1977)Cohen-Tannoudji, Diu, and
  Laloë]{Cohen-Tannoudji:101367}
Claude Cohen-Tannoudji, Bernard Diu, and Franck Laloë.
\newblock \emph{{Quantum mechanics; 1st ed.}}
\newblock Wiley, New York, NY, 1977.

\bibitem[Potters and Bouchaud(2020)]{potters_bouchaud_2020}
Marc Potters and Jean-Philippe Bouchaud.
\newblock \emph{A First Course in Random Matrix Theory: for Physicists,
  Engineers and Data Scientists}.
\newblock Cambridge University Press, 2020.

\bibitem[Jackson(1975)]{Jackson:100964}
John~David Jackson.
\newblock \emph{{Classical electrodynamics; 2nd ed.}}
\newblock Wiley, New York, NY, 1975.

\bibitem[Ledoit and P{\'e}ch{\'e}(2011)]{ledoit2009eigenvectors}
Olivier Ledoit and Sandrine P{\'e}ch{\'e}.
\newblock Eigenvectors of some large sample covariance matrix ensembles.
\newblock \emph{Probability Theory and Related Fields}, 151\penalty0
  (1):\penalty0 233--264, 2011.

\bibitem[Benaych-Georges(2009)]{Benaych-Georges2009}
Florent Benaych-Georges.
\newblock Rectangular random matrices, related convolution.
\newblock \emph{Probability Theory and Related Fields}, 144\penalty0
  (3):\penalty0 471--515, 2009.

\bibitem[Pennington and Worah(2017)]{NIPS2017_0f3d014e}
Jeffrey Pennington and Pratik Worah.
\newblock Nonlinear random matrix theory for deep learning.
\newblock In I.~Guyon, U.~V. Luxburg, S.~Bengio, H.~Wallach, R.~Fergus,
  S.~Vishwanathan, and R.~Garnett, editors, \emph{Advances in Neural
  Information Processing Systems}, volume~30, 2017.

\bibitem[Louart et~al.(2018)Louart, Liao, and Couillet]{Louart}
Cosme Louart, Zhenyu Liao, and Romain Couillet.
\newblock A random matrix approach to neural networks.
\newblock \emph{The Annals of Applied Probability}, 28\penalty0 (2):\penalty0
  1190--1248, 2018.

\bibitem[Menon(2017)]{menon2017complex}
Govind Menon.
\newblock The complex {B}urgers’ equation, the {HCIZ} integral and the
  {C}alogero-{M}oser system.
\newblock \emph{Preprint}, 2017.

\bibitem[Dupic and Castillo(2014)]{dupic2014spectral}
Thomas Dupic and Isaac~P{\'e}rez Castillo.
\newblock Spectral density of products of {W}ishart dilute random matrices.
  part {I}: the dense case.
\newblock \emph{Preprint arXiv:1401.7802}, 2014.

\bibitem[Sanov(1958)]{sanov1958probability}
Ivan~N Sanov.
\newblock \emph{On the probability of large deviations of random variables}.
\newblock United States Air Force, Office of Scientific Research, 1958.

\end{thebibliography}

\appendix

\section{Details on the regularization of logarithmic
integrals}\label{app.regularization}

In the expression for the free entropy \cref{eq.freeentropy} and for the MMSE
\cref{eq.res2}, we have integrals of the form
\begin{eqnarray}
       \dashint d\lambda \, d\zeta \,  \denshat{\bY}(\lambda)
        \denshat{\bY}(\zeta) \log|\lambda-\zeta| \, , & \dashint d\lambda \,
        \denshat{\bY}(\lambda) \log|\lambda| \, ,
\end{eqnarray}
which have logarithmic divergences, respectively, on the diagonal $\zeta =
\lambda$ and at the origin $\lambda = 0$. This integrals must be intended as
continuum limits of the corresponding discrete expressions for finite sized
matrices, i.e.
\begin{equation}
    \dashint d\lambda \, d\zeta \,  \denshat{\bY}(\lambda) \denshat{\bY}(\zeta) \log|\lambda-\zeta|
    = \lim_{N \to \infty} \frac{1}{N^2} \sum_{i, j = 1}^N \log|\sigma_i-\sigma_j|
    \, ,
\end{equation}
where $\{\sigma_i\}$ is the singular spectrum of a size $N$ matrix whose
limiting singular value density converges to $\denshat{\bY}$, and
\begin{equation}
    \dashint d\lambda \, \denshat{\bY}(\lambda) \log|\lambda|
    = \lim_{N \to \infty} \frac{1}{N^2} \sum_{\substack{i = 1\\\sigma_i \neq 0}}^N \log|\sigma_i|
    \, .
\end{equation}
As the spacing between singular values goes to zero as $N\to\infty$, the correct
way to interpret the integrals is, respectively,
\begin{equation}
    \dashint d\lambda \, d\zeta \,  \denshat{\bY}(\lambda) \denshat{\bY}(\zeta) \log|\lambda-\zeta|
    = \lim_{\epsilon \to 0}
    \int d\lambda \, d\zeta \,  \denshat{\bY}(\lambda) \denshat{\bY}(\zeta) \log|\lambda-\zeta| \, \mathbb{I}(|\zeta - \lambda| > \epsilon)
    \, ,
\end{equation}
where $\mathbb{I}(\cdot)$ is the indicator function, and
\begin{equation}
    \dashint d\lambda \, \denshat{\bY}(\lambda) \log|\lambda|
    = \lim_{\epsilon \to 0}
    \int d\lambda \, \denshat{\bY}(\lambda) \log|\lambda|
    \,
    \mathbb{I}(|\lambda| > \epsilon)
    \, .
\end{equation}

\section{Details of numerical integration of \cref{eq.res2}}\label{sec.tricks}

The only numerical difficulty of the paper is the estimation of the integrals in
\cref{eq.res2} in the case of the factorized prior \cref{eq.defWishart}. The
integral is performed by quadrature using Simpson's rule. The function we
integrate has a wide support and it peaks around the origin for $R_2<1$, so we
must have more integration points in this region.

We are left with two sub-problems: estimating the support of $\denshat{\bY}$,
and the support of the peak at the origin. The edges of the support of the
signal $\Sstar$ can be computed exactly following \cite{dupic2014spectral}, and
in particular their Appendix B, with the parameter conversions
\begin{equation}
    \alpha_1 = R_2 \, , \alpha_2 = R_1 \, 
\end{equation}
giving a bound $\sigma_{\rm max}(\Sstar)\leq\sigma_{\rm signal}$. An upper-bound
on the largest singular value of the full observation $\bY$ can be obtained by
summing the largest support edge computed above and the largest support edge of
the Marchenko-Pastur distribution accounting the singular value density of the
noise,

\begin{equation}
    \sigma_{\rm max}(\sqrt{\Delta} \bZ) \leq \frac{\Delta (1 + \sqrt{R_1})^2}{\sqrt{R_1}} = \sigma_{\rm noise}\,.
\end{equation}
Putting everything together we get:
\begin{equation}
    \sigma_{\rm max}(\bY) \leq \sigma_{\rm signal} + \sigma_{\rm noise} \, .
\end{equation}
The right-most edge of the peak at the origin can be upper bounded empirically
by $\sigma_{\rm noise}$. While there are surely tighter bounds, this one allows
for reasonable performance in the integration. 

\section{The MMSE of low-rank denoising}\label{app.lowrank}

The spiked matrix denoising problem is studied in \cite{2017Thibault,
miolane2018fundamental}. We are given an observation matrix of the form:

\begin{equation}
    \bY = \frac{uv^T}{\sqrt{m}} + \sqrt{\Delta} \bZ
\end{equation}
where $u \in \bbR^{m}$, $v \in \bbR^{p}$ and $\bZ \in \bbR^{m\times p}$ all have
standard Gaussian entries. The MMSE for our problem takes the form \cite[section
2.6]{miolane2018fundamental}:
\begin{equation}
    \MMSE_{\rm spiked} = 1 - q_u q_v,
\end{equation}
where $q_u$ and $q_v$ are solutions of the fixed point equations:

\begin{equation}
\begin{cases}
q_u = F\left(\frac{R_1q_v}{\Delta}\right),\\
q_v =  F\left(\frac{q_u}{\Delta}\right)\, ,
\end{cases} 
\end{equation}
with
\begin{equation}
    F(q) = \EE_{x_0^*,z \sim \mathcal{N}(0,1)} \left[ \frac{\int x_0^*x_0e^{x_0 z\sqrt{q} + x_0 x_0^*q -
    \frac{1}{2}x_0^2q}dP(x_0)}{\int e^{x_0 z\sqrt{q} + x_0 x_0^*q -
    \frac{1}{2}x_0^2q}dP(x_0)} \right],
\end{equation}
where $P(x)$ is the standard Gaussian PDF. The integral above can be evaluated
explicitly, yielding:

\begin{equation}
\begin{cases}
q_u = \frac{R_1 q_v}{\Delta + R_1 q_v}\, ,\\
q_v = \frac{q_u}{\Delta + q_u}\, ,\\
\end{cases} 
\end{equation}
whose solution is
\begin{equation}
\begin{cases}
q_u = \frac{R_1 - \Delta^2}{\Delta + R_1}\, ,\quad
q_v = \frac{R_1 - \Delta^2}{(\Delta + 1) R_1} \quad &\text{for} \quad 0 < \Delta < \sqrt{R_1} \, ,\\
q_u = q_v = 0 \quad &\text{for} \quad \Delta > \sqrt{R_1} \, .\\
\end{cases} 
\end{equation}
There is an undetectable phase for $\Delta>\sqrt{R_1}$. In the main text the
observation matrix is normalized differently, which is why in
Fig.~\ref{fig_low_rank} we have the undetectable phase for $\Delta
R_2>\sqrt{R_1}$

\section{The I-MMSE theorem}\label{app.I-MMSE}

Let us prove Eq.~\eqref{eq.mmsePHI}, i.e.\ that 
\begin{equation}
    \MMSE = \EE_{\Sstar, \bY}\left[ \MSE\left(\Sstar, 
    \left\langle \bS \right\rangle_{\bY} \right) \right]
    = \Delta + 2 \Delta^2 \del_{\Delta} \EE\left[ \Phi_{\bY} \right]
    \, ,
\end{equation}
where in the following $\EE_{\Sstar, \bY}$ denotes a joint average over $\Sstar$
and $\bY$, and $\left\langle \cdot \right\rangle_{\bY}$ the posterior average at
fixed $\bY$. We start by expressing the MMSE more explicitly
\begin{equation}
    \begin{split}
      \MMSE &= \frac{1}{\sqrt{mp}} \EE_{\Sstar, \bY}||\Sstar - \left\langle \bS \right\rangle_{\bY} ||_2^2
      \\
        &=\frac{1}{\sqrt{mp}}  \EE_{\Sstar, \bY}\left[||\Sstar||_2^2 
            +  || \left\langle \bS \right\rangle_{\bY} ||_2^2
            - 2  \Tr\left(\Sstar  \left\langle \bS \right\rangle_{\bY}^T \right)
            \right] \,. 
    \end{split}
\end{equation}
Now, we use Nishimori's identity \cite{Nishimori_1980} on the last term, i.e.
\begin{equation}
    \EE_{\Sstar, \bY}[\left\langle g(\Sstar,  \bS ) \right\rangle_{\bY} ]
    =
    \EE_{\bY}[\left\langle g(\bS_1,  \bS_2 ) \right\rangle_{\bY} ]
    \, ,
\end{equation}
where $\bS_{1, 2}$ are two independent random variables distributed with the
posterior distribution, and $g$ a continuous bounded function. In words, under
Bayes optimality, the ground-truth is indistinguishable from a sample from the
posterior for what concerns averages. In our particular case, 
\begin{equation}
    \EE_{\Sstar, \bY} \left[ \Tr\left(\Sstar  \left\langle \bS \right\rangle_{\bY}^T \right) \right]
    = 
    \EE_{\bY} \left[ \Tr\left(\left\langle \bS \right\rangle_{\bY} \left\langle \bS \right\rangle_{\bY}^T \right) \right]
    = 
    \EE_{\bY} || \left\langle \bS \right\rangle_{\bY} ||_2^2 
    \, .
\end{equation}
Thus
\begin{equation}
    \begin{split}
        \EE_{\Sstar, \bY} \left[ \MSE\left(\Sstar, \left\langle \bS \right\rangle_{\bY} \right) \right]
        &= \frac{1}{\sqrt{mp}} \EE_{\Sstar, \bY}\left[ 
            ||\Sstar||_2^2 -  || \left\langle \bS \right\rangle_{\bY} ||_2^2
            \right] \,. 
    \end{split}
\end{equation}
The second step is to compute $\EE_{\Sstar,
\bY}\left[\del_{\Delta^{-1}}\Phi_{\bY}\right]$
\begin{equation}
    \begin{split}
      \EE_{\Sstar, \bY}\left[\del_{\Delta^{-1}}\Phi_{\bY}\right] 
        &=
        \frac{1}{mp} \EE_{\Sstar, \bY}\left[\frac{1}{\caZ_{\bY}}\del_{\Delta^{-1}}\caZ_{\bY}\right] 
         \\ &=
        \frac{\Delta}{2}
        - \frac{1}{2\sqrt{mp}}
        \EE_{\Sstar, \bY}\left[
            ||\Sstar||_2^2 -2 ||\left\langle \bS \right\rangle_{\bY}||_2^2 + \left\langle || \bS ||_2^2 \right\rangle_{\bY}
        -   \sqrt\Delta \sum_{i\mu}  \bZ_{i\mu} \left\langle \bS_{i\mu} \right\rangle_{\bY} \right] 
    \end{split}
\end{equation}
where we used that $\EE_{\bZ} \bZ = 0$, and again Nishimori's identity. Then,
one notices that
\begin{equation}
  \EE_{\Sstar, \bY}\left[ \sqrt\Delta \sum_{i\mu} \left\langle_{\bY}  \bZ_{i\mu}\bS_{i\mu} \right\rangle_{\bY} \right]
    =
    \EE_{\Sstar, \bY}\left[  \left\langle || \bS ||_2^2 \right\rangle_{\bY} - ||\left\langle \bS \right\rangle_{\bY}||_2^2 \right] 
\end{equation}
by using Stein's lemma
\begin{equation}
    \EE_{\Sstar, \bY}\left[ \sqrt\Delta \sum_{i\mu}  \bZ_{i\mu} \left\langle_{\bY}  \bS_{i\mu} \right\rangle \right]
    =
    \EE_{\Sstar, \bY}\left[ \sqrt\Delta \sqrt[4]{mp} \sum_{i\mu} \frac{\del \left\langle \bS_{i\mu} \right\rangle_{\bY}}{\del  \bZ_{i\mu}}  \right]
\end{equation}
and then tediously computing the derivative. Finally 
\begin{equation}
    \begin{split}
      \EE_{\bY}\left[\del_{\Delta^{-1}}\Phi_{\bY}\right] 
        =
        \frac{\Delta}{2}
        - \frac{1}{2\sqrt{mp}}
        \EE_{\Sstar, \bY} \left[
            ||\Sstar||_2^2 - ||\left\langle \bS \right\rangle_{\bY}||_2^2  \right] 
        = \frac{\Delta}{2} - \frac{1}{2} \MMSE 
    \end{split}
\end{equation}
as anticipated.

\section{The high-dimensional limit of \cref{eq:ZY}}\label{app.sp}

We shortly justify in this part why the high-dimensional limit of \cref{eq:ZY}
indeed yields \cref{eq.phiprehciz}. This is actually the consequence of a more
general result, that we describe here. Let us consider an expression of the type
$G = \lim_{m \to \infty} G_m$, with
\begin{equation}\label{eq:Gm_1}
    G_m \equiv \frac{1}{m^2} \ln \int_{\bbR^m} \rd \bT \, \exp\{m^2 F(\mu_T^{m})\},
\end{equation}
in which $\mu_T^m$ is the empirical distribution of $\bT$, i.e.\ 
\begin{equation}
    \mu_T^m \equiv \frac{1}{m} \sum_{\mu=1}^m \delta_{T_i}.
\end{equation}
In particular, one can check easily that \cref{eq:ZY} is exactly of this type
(it is easy to see that all terms inside the integral on $\bT$ are of order
$\exp\{\Theta(m^2)\}$, note that for the HCIZ integral this is a consequence of
the analysis of \cite{guionnet2021large}. In \cref{eq:Gm_1} one can use Sanov's
theorem \cite{sanov1958probability}, i.e.\ a large deviations principle on
$\mu_T^m$. Informally, Sanov's theorem states that the law of the random measure
$\mu_T^m$ satisfies: 
\begin{equation}\label{eq:sanov}
    \frac{1}{m} \ln \bbP[\mu_T^m \simeq \mu] \simeq - \int \mu(\rd x) \ln (\rd \mu / \rd x),
\end{equation}
in which the right hand side is (minus) the relative entropy with respect to the
Lebesgue measure. The crucial point is that \cref{eq:sanov} implies that
$\bbP[\mu_T^m \simeq \mu]$ is of typical size $\exp\{\Theta(m)\}$, while in
\cref{eq:Gm_1} we consider asymptotics in the scale $\exp\{\Theta(m^2)\}$! In
particular, we can apply the Laplace method in this scale on the empirical
distribution $\mu_T^m$, and the entropic term of \cref{eq:sanov} will not
contribute. Therefore we reach:
\begin{equation}
    G = \lim_{m \to \infty} G_m = \sup_{\mu} F(\mu),
\end{equation}
which is what we wanted to show. In order to reach \cref{eq.phiprehciz}, there
only remains to argue that one can exactly work out this supremum using
Nishimori's identity, cf.\ our remark above \cref{eq.phiprehciz}.

\section{The Hellman-Feynman theorem adapted to singular values}
\label{app.HF}

Let us consider a rectangular matrix $\bM \in \bbR^{m \times p}$ with real
non-negative singular values $\{y_l\}_{l=1}^{m}$ and orthonormal singular
vectors $\{\vu_l\}_{l=1}^{m}$ (left) and $\{\vv_l\}_{l=1}^{m}$ (right). Singular
values act as if they where eigenvalues. Explicitly, 
\begin{equation}\label{eq.svrel}
    \bM \cdot \vv_l = y_l \vu_l \, , \quad
    \vu_l^T \cdot \bM  = y_l \vv_l^T \quad \text{ and } \quad
    y_l = \vu_l^T \cdot \bM \cdot \vv_l
    \, .
\end{equation} 
Notice that the right singular vectors must be completed by $p - m$ other
orthogonal vectors in order to form a basis for the domain of $\bM$.

We consider now a matrix $\bM(\lambda)$ that depends on a parameter $\lambda$.
Consequently, both eigenvalues and singular values will depend on $\lambda$, and
we would like to compute the derivative w.r.t. $\lambda$ of the singular values.
We restrict to the case of non-degenerate singular spectrum.

We fix a particular singular value $y_l$ with singular vectors $\vu_l$ and
$\vv_l$ (we drop the $l$ subscript in the following, as well as the dependence
on $\lambda$). We have that
\begin{equation}
    \begin{split}
        \del_\lambda y 
        &= 
        \del_\lambda \left( \vu^T \cdot \bM \cdot \vv \right)
        = 
        \vu^T \cdot \bM  \cdot \del_\lambda \left(  \vv \right) +
        \vu^T \cdot \del_\lambda \left( \bM \right) \cdot \vv +
        \del_\lambda \left( \vu^T  \right) \cdot \bM \cdot \vv
        \\
        &= 
        y \, \vv^T \cdot \del_\lambda \left(  \vv \right) +
        \vu^T \cdot \del_\lambda \left( \bM \right) \cdot \vv +
        y \, \del_\lambda \left( \vu^T  \right) \vu
        \\
        &= 
        \frac{y}{2}  \del_\lambda \left(  \vv^T \cdot \vv  +  \vu^T \cdot \vu \right) +
        \vu^T \cdot \del_\lambda \left( \bM \right) \cdot \vv 
        \\ 
        &= 
        \vu^T \cdot \del_\lambda \left( \bM \right) \cdot \vv 
        \, .
    \end{split}
\end{equation}
where we used the relations \cref{eq.svrel} and the fact that $\vu$ and $\vv$
are normalized. Thus
\begin{equation}
    \del_\lambda y = \vu^T \cdot \del_\lambda \left( \bM \right) \cdot \vv 
    \, .
\end{equation}
The original version of the Hellman-Feynman theorem for symmetric matrices has a
very similar proof. In that case, left and right singular vector coincide.

\section{Details of \cite{NIPS2017_0f3d014e}}
\label{app.proofPennington} 

\subsection{Extension of the proof to random non-linearities}

The proof strategy of \cite{NIPS2017_0f3d014e} --- based on the method of
moments --- is presented in \cite[Supplementary Material, Section
1.2]{NIPS2017_0f3d014e}. We would like to extend it to the case of random
non-linearities. The main observation is that no hypothesis on the non-linearity
$f$ is ever done in the proof up until Eq. (S27). Before that, the logic
presented by the authors holds \textit{as is} for random $f$, provided that an
average over $f$ is added in Eq. (S2).

Thus, to incorporate random non-linearities into the proof of
\cite{NIPS2017_0f3d014e} one just needs to analyze Eq. (S27) and Eq. (S29)
therein. Notice that in the random cases the first passage of both equations is
the same, with the addition of an average over the randomness of $f$ to be
performed last. In both cases, the crucial passage is the fact that the
non-linearity is \textit{factorized} over the coordinates of the matrices. In
order for this factorization to hold in the random case, we just need to require
that the randomness introduced by $f$ is i.i.d.\ over coordinates, so that the
average over $f$ factorizes as well.

Thus, for non-linearities with i.i.d.\ randomness the proof holds as is, and the
effect of the randomness of $f$ will be just a modification to the definition of
the parameters $\eta$ and $\zeta$, which becomes
\begin{equation}
    \eta = \EE_f \int Dz f(z)^2 \quad\text{ and }\quad \zeta = \EE_f \left[\int Dz f'(z)\right]^2 \, ,
\end{equation}
where $\EE_f$ denotes averaging with respect to the randomness of $f$. In the
case presented in this paper $\eta = (1+\Delta)\sqrt{R_1}$, $\zeta = \sqrt{R1}$. 

\subsection{Stieltjes transform}

The Stieltjes transform of the symmetrized singular values $g_\bY(z)$ for
factorized priors can be written using \cref{eq.relG} as $zG_z$, where $G_z$ is
a root of equation $\sum_{k=0}^4 a_k(z) G_z^k = 0$ with coefficients:
\begin{equation}\label{eq.quartic}
\begin{split}
&a_0 = -\psi^3 \\
&a_1 = \psi ( \zeta (\psi - \phi) + \psi ( \eta (\phi - \psi) + \psi z^2 ) ) \\
&a_2 = - \zeta^2  (\phi - \psi)^2 + \zeta ( \eta (\phi - \psi)^2 + \psi z^2 (2 \phi - \psi) ) - \eta \psi^2 z^2 \phi \\
&a_3 = - \zeta z^2 \phi ( 2 \zeta \psi - 2 \zeta \phi - 2 \eta \psi + 2 \eta \phi + \psi z^2 ) \\
&a_4 = \zeta z^4 \phi^2 (\eta - \zeta) \, , 
\end{split}
\end{equation}
where $\phi = R_2/R_1$ and $\psi = R_2$.



\section{Tools to compute the derivative of \cref{eq.Gt}}\label{app.hciz}

To show that $G(t)$ defined in \cref{eq.Gt} is a constant function of $t$ one
needs two technical bits. The first one is that \cite[lemma
C.1]{maillard2021perturbative}:
\begin{equation}\label{eq.rhoCubed}
    \frac{1}{3} \int dx \, \hrho(x)^3 = \int dx \, \hrho(x) \left( \mathcal{H}[\hrho](x) \right)^2 \, ,
\end{equation}
and the second one is that \cite[see between Eq. 4.20 and
4.21]{guionnet2021large}
\begin{equation}
    \int dx \, \hrho_t (x) \mathcal{H}[\hrho_t](x) \frac{1}{x} = 0 \, ,
\end{equation}
due to the symmetry of all symmetrized singular value densities. Using these two
properties one can directly compute the time derivative of \cref{eq.Gt} and
simplify all terms, showing that the derivative is null.

\end{document}